\documentclass[smallextended]{svjour3}
\usepackage{graphicx}
\usepackage{natbib}
\usepackage{txfonts}
\begin{document}
\title{On type-I migration near opacity transitions} 
\subtitle{A generalized
  Lindblad torque formula for planetary population synthesis}

\titlerunning{Type~I migration at opacity transitions}
\author{Fr\'ed\'eric S. Masset}

\institute{Fr\'ed\'eric Masset \at
             Institute of Physical Sciences\\
             Universidad Nacional Aut\'onoma de M\'exico\\
             Apdo Postal 48--3\\
             62251 Cuernavaca, Mor.\\
             Mexico\\
            \email{masset@fis.unam.mx} } 
\maketitle

\begin{abstract}
  We give an expression for the Lindblad torque acting on a low-mass
  planet embedded in a protoplanetary disk that is valid even at
  locations where the surface density or temperature profile cannot be
  approximated by a power law, such as an opacity transition. At such
  locations, the Lindblad torque is known to suffer strong deviation
  from its standard value, with potentially important implications for
  type~I migration, but the full treatment of the tidal interaction is
  cumbersome and not well suited to models of planetary population
  synthesis. The expression that we propose retains the simplicity of
  the standard Lindblad torque formula and gives results that
  accurately reproduce those of numerical simulations, even at
  locations where the disk temperature undergoes abrupt changes. Our
  study is conducted by means of customized numerical simulations in
  the low-mass regime, in locally isothermal disks, and compared
  to linear torque estimates obtained by summing fully analytic
  torque estimates at each Lindblad resonance. The functional
  dependence of our modified Lindblad torque expression is suggested
  by an estimate of the shift of the Lindblad resonances that mostly
  contribute to the torque, in a disk with sharp gradients of
  temperature or surface density, while the numerical coefficients of
  the new terms are adjusted to seek agreement with numerics. As side
  results, we find that the vortensity related corotation torque
  undergoes a boost at an opacity transition that can counteract
  migration, and we find evidence from numerical simulations that
  the linear corotation torque has a non-negligible dependency upon
  the temperature gradient, in a locally isothermal disk.

\end{abstract}

\section{Introduction}
\label{sec:introduction}
As the number of extrasolar planets discovered increases continuously,
there is a growing volume of models of planet population synthesis, in
which one evolves a collection of planets in a model of protoplanetary
disk, so as to compare the statistics of real planetary systems to
those of the model outcomes
\citep{2010ApJ...715L..68L,Mordasini:2009p398,Mordasini:2009p399,Ida:2008p297,Ida:2008p396}.
One key feature of models of planetary population synthesis is the
tidal torque that exists between a planet and its parent
protoplanetary disk, and which drives a radial migration of the
planet. Since this torque is large, planets migrate a substantial
fraction of their initial semi major axis over the disk lifetime, and
a model's outcome strongly depends on the underlying torque
expression.  The tidal torque has two components:
\begin{itemize}
\item The differential Lindblad torque, exerted by a wake excited by
  the planet in the neighboring flow, where the latter becomes
  supersonic with respect to the planet, owing to the strong Keplerian
  shear.
\item The corotation torque, which corresponds to an exchange of
  angular momentum between the planet and the coorbital material.
\end{itemize}
Early efforts have intensively focused on the differential Lindblad
torque, which is now known in great detail in power law disks. The
well known standard formula of \citet{tanaka2002} has proved to
account for the torque measured in numerical simulations with
remarkable accuracy. More recently, the corotation torque has received
a detailed attention, and was shown to have two components: one
related to the vortensity\footnote{The vortensity is here defined as
  the ratio of the vertical component of the flow vorticity over the
  disk's surface density.} gradient across the orbit, and another one
related to the entropy gradient across the
orbit \citep{bm08,pp08,cm09,mc09,2010ApJ...723.1393M,pbck10}. The
corotation torque, moreover, depends on the dissipative processes at
work in the disk. \citet{2010ApJ...723.1393M} have provided general
corotation torque formulae that may be applied to planets embedded in
disks with arbitrary vortensity and entropy profiles, and with
arbitrary values of the viscosity and thermal diffusivity.

The domain of validity of the formulae for the corotation torque now
supersedes that of the Lindblad torque: the corotation torque formulae
apply to arbitrary profiles, and only requires the evaluation of
radial derivatives at the orbit, whereas the Lindblad torque standard
formula exclusively applies to disks where the surface density and
temperature profiles are power laws of the radius. When this is not
the case, one has to resort to basics and evaluate the torque
individually at each Lindblad resonance. This was the approach
undertaken by \citet{mg2004}, who stressed the high sensitivity of the
torque to the resonance location. Their key result was that the
Lindblad torque can be substantially reduced at locations, in the
disk, where the local profiles of surface density and/or temperature
cannot be approximated by power laws, namely at opacity transitions.

This approach, nonetheless, would be very costly in planetary
population synthesis models, as it would require the evaluation of a
series at each time step, for each planet. To the best of our
knowledge, all the planetary population synthesis models performed
thus far have made use, in different flavors, of standard torque
formulae for type~I migrating objects, in some case including a
uniform, {\em ad hoc} reduction factor so as to moderate the drastic
effects of standard type~I migration.

The present work is an attempt to provide a simple generalization of
the standard Lindblad torque formula, that can be applied in arbitrary
disk models ({\em i. e.} where the profiles of surface density and
temperature are not power laws of radius). The reader may glimpse at
Eq.~(\ref{eq:79}), which constitutes our main result, and compare it
with Eq.~(\ref{eq:10}), which is the standard expression in power law
disks. As can be seen, there is very little additional complexity in
evaluating Eq.~(\ref{eq:79}), which can be readily implemented in
models of planetary population synthesis.

We firstly define the notation in section~\ref{sec:notation}.{We then
  discuss in some detail the relevance of numerical simulations in the
  low-mass regime in assessing the linear torques in
  section~\ref{sec:relev-numer-simul}}, and we present standard torque
expressions in section~\ref{sec:stand-lindbl-torq}. In particular, we
need to calibrate the corotation torque in order to properly subtract
it from our runs, so that we get exclusively the Lindblad torque. We
present our method in section~\ref{sec:assumptions-set-up}. We then
derive the effects of higher order derivatives of the surface density
or temperature on the Lindblad torque in
section~\ref{sec:indiv-effects-high}. As we shall see, the end result
is particularly simple, as it turns out that only the third derivative
of the temperature has a significant impact on the Lindblad torque. We
then apply this formula to a disk model aimed at representing an
opacity transition in section~\ref{sec:torque-behavior-at}, where we
present the generalized torque expression in definitive form. In that
section we also estimate the corotation torque. We find that at the
transition the corotation torque also features a term proportional to
the third derivative of the temperature, with a sign opposite to that
of the Lindblad torque.  We discuss our results in
section~\ref{sec:discussion} and draw our conclusions in
section~\ref{sec:conclusions}.

\section{Notation}
\label{sec:notation} 
We consider a planet of mass $M_p$ orbiting a star of mass $M_*$ on a
circular orbit of radius $a$, with orbital frequency $\Omega_p$. The
planet is embedded in a gaseous disk with surface density $\Sigma(r)$
and temperature $T(r)$. The vertical scale height $H(r)$ of the disk is
related to the gas sound speed $c_s(r)$ and to the local Keplerian
frequency $\Omega_K(r)$ by:
\begin{equation}
  \label{eq:1}
  H(r) = \frac{c_s(r)}{\Omega_K(r)}.
\end{equation}
We oftentimes use the disk aspect ratio or relative thickness defined
as $h(r)=H(r)/r$, and the planet to star mass ratio $q=M_p/M_*$. We denote
$\omega$ the vertical component of the flow's vorticity:
\begin{equation}
\label{eq:2}
\omega=\frac{1}{r}\partial_r(r^2\Omega).
\end{equation}

We define hereafter notation specific to non power law disks.  We
denote with $\alpha_i$ a quantity related to the $i^{th}$ derivative
of the surface density:
\begin{equation}
  \label{eq:3}
  \alpha_i = (-1)^{\min(i,2)}h^{i-1}\frac{d^i\log\Sigma_0}{(d\log r)^i}
\end{equation}
We thus have:
\begin{equation}
  \label{eq:4}
  \alpha_1 = -\frac{d\log\Sigma}{d\log r},
\end{equation}
\begin{equation}
  \label{eq:5}
  \alpha_2 = h\frac{d^2\log\Sigma}{d\log r^2},
\end{equation}
\begin{equation}
  \label{eq:6}
  \alpha_3 = h^2\frac{d^3\log\Sigma}{d\log r^3}.
\end{equation}
In a power law disk, all $\alpha_i$ with $i > 1$ cancel out and we
have $\Sigma_0 \propto r^{-\alpha_1}$. Our value of  $\alpha_1$ coincides
with the standard notation $\alpha$ of \citet{tanaka2002}. The
factor $(-1)^{\min(i,2)}$ in Eq.~(\ref{eq:3}) is exclusively meant
to conserve the minus sign for the first derivative, so as to save the
standard practice. In contrast, higher order derivatives do not
feature this minus sign. In a similar fashion we denote with $\beta_i$
a quantity related to the $i^{th}$ derivative of the temperature:
\begin{equation}
  \label{eq:7}
  \beta_i= (-1)^{\min(i,2)}h^{i-1}\frac{d^i\log T_0}{(d\log r)^i}.
\end{equation}
Our $\beta_1$ therefore coincides with the value of $\zeta$ of
\citet{2010ApJ...724..730D}, or with the value of $\beta$ of
\citet{pbck10}, but it is twice the value of $\beta$ of
\citet{tanaka2002}, where $\beta$ is the slope of the sound speed
rather than the temperature.  We work out how the differential
Lindblad torque behaves in a disk in which some of the $\alpha_i$
and/or $\beta_i$ ($i>1$) do not cancel out. We limit ourselves to $i
\leq 3$. As we shall see in section~\ref{sec:torque-behavior-at},
there is no need to push the expansion further. We entertain
separately the role of the $\alpha_i$ and $\beta_i$ ($i >1$), assuming
that the resonance shift is sufficiently small that the global effect
is obtained by adding up individual contributions of higher
derivatives of surface density and temperature. We shall often refer
to the residue of the surface density profile with respect to the
power law surface density with same $\alpha_1$ as a perturbation (the
same holds for perturbations of temperature). It should be noted
however that this perturbation is not related to the planet. It is
axisymmetric, and the ``perturbed'' disk is in rotational equilibrium.

\section{Relevance of numerical simulations in assessing the torques
  acting on a low-mass planet}
\label{sec:relev-numer-simul}

Since full hydrodynamical simulations are non-linear in nature, it is
questionable whether the torques obtained in simulations in
  the low-mass regime can be regarded as estimates of the linear
  torque value. As a prerequisite, it is therefore necessary to
confront the torque thus obtained to the torque obtained independently
by a linear analysis. This is the work of
\citet{2009arXiv0901.2265P}. Part of their findings (the ones that we
shall need in this work) are as follows:
  \begin{enumerate}
  \item For low-mass planets, non-linear effects are found at all
    masses, and identified with the onset of the horseshoe drag
    regime, provided the disk's viscosity is sufficiently low. In that
    case, the linear corotation torque exists only temporarily upon
    insertion of the planet. The timescale for the onset of non-linear
    effects scales with $q^{-1/2}$.
  \item In the inviscid case, the (non-linear) corotation torque
    saturates at larger time, leaving only the linear Lindblad torque.
  \item At early time or at any time for large viscosity, the total
    torque acting on the planet is found to be in excellent agreement
    with the total linear torque.
  \end{enumerate}
  \citet{2009arXiv0901.2265P} established these results in disks with
  a locally isothermal equation of state, that is disks in which the
  sound speed or temperature is an arbitrary function of $r$, constant
  in time. Admittedly, they obtained these results by considering
  various temperature and surface density profiles which are
  exclusively power laws of the radius.

  We shall use the result~2 above in the very same framework (see
  appendix~\ref{sec:calibr-lindbl-torq}), {\em i. e.} for power law
  disks, in order to establish an expression for the Lindblad torque
  in these disks, that is valid for our particular setup (two
  dimensional disks and a planetary potential softened over a given
  length.) In regard to the above we shall consider that expression as
  a good approximation of the linear Lindblad torque.

  We shall use the result~3 also in power law disks (see
  appendix~\ref{sec:calibr-lindbl-torq}), so as to get an expression
  for the total torque, which, again, we regard as a good
  approximation of the total linear torque for the reasons listed
  above. This enables us to get an expression of the corotation torque
  at early times (by subtracting the Lindblad torque), as a function
  of the gradients of vortensity and temperature. This step is needed
  to enable us to subtract the corotation torque from the runs in
  which the disk profiles are not power laws of radius, so as to get
  the Lindblad torque.

  Finally, we shall measure the total torque at early time in
  calculations where the surface density or temperature profiles are
  not power laws of the radius. Although, strictly speaking, this is
  not contemplated in the framework of \citet{2009arXiv0901.2265P}, it
  is reasonable to consider that the torque acting on the planet has a
  linear value at this stage. Regardless of the profiles, the response
  of a disk to a planet of low mass ($q/h^3 \ll 1$) should be linear
  over a time interval that can be made arbitrarily large by
  decreasing the planetary mass. We further comment that in our case,
  we have $q/h^3=1.6\cdot 10^{-2}$, almost an order of magnitude
  smaller than that of the fiducial calculation of
  \citet{2009arXiv0901.2265P}, which is $q/h^3=0.1$.

  We conclude that numerical simulations, albeit non-linear in
  essence, can be used to assess with a good accuracy the linear
  torques acting on a low-mass planet.  The degree of accuracy
    can be estimated either by comparing our data to the expression of
    \citet{pbck10} for power law profiles, or by evaluating the
    fitting error of our data (see
    appendix~\ref{sec:calibr-lindbl-torq}). From both methods one can
    infer a degree of accuracy of a few percents, which is far enough
    for our purpose.

\section{Standard torque expressions}
\label{sec:stand-lindbl-torq}
Standard Lindblad torque formulae, be they two or three dimensional,
all share the same form\footnote{As a convention throughout this work,
  all torques refer to the torque exerted by the disk onto the
  planet.}:
\begin{equation}
  \label{eq:8}
  \Gamma_L = -(k_0+k_1\alpha_1+k_2\beta_1)\Gamma_{\rm ref},
\end{equation}
where $k_0$, $k_1$ and $k_2$ are numerical constants of order unity,
and where $\Gamma_{\rm ref}$ is the torque normalization factor given
by:
\begin{equation}
  \label{eq:9}
  \Gamma_{\rm ref} = \Sigma a^4\Omega_p^2q^2h^{-2}.
\end{equation}
We refer to Lindblad torque formulae having the form of
Eq.~(\ref{eq:8}) as {\em standard} to express that they exclusively
involve a linear combination of the surface density and temperature
first derivatives at the position of the orbit. Strictly speaking,
standard Lindblad torque formulae are valid only in disks in which
both surface density and temperature are power laws of the radius.

The expression of the Lindblad torques alone are not explicitly given
by \citet{tanaka2002}, but they can be inferred from their Tab.~1
and~2 \citep[see][]{2008EAS....29..165M}. This yields
$(k_0,k_1)=(3.2,1.468)$ in the two dimensional case without softening,
and $(k_0,k_1)=(2.34,-0.099)$ in the three dimensional case. Note that
$k_2$ was left unexplored in this original work, as the authors
restricted themselves to globally isothermal disks ({\em i.e.} without
a temperature gradient).

In the present work we shall deal with two dimensional disks and a
softened planetary potential (we chose a softening length of the
potential $\epsilon=0.6H$). For our purpose, we therefore have to
firstly establish the standard Lindblad torque formulae in those
disks. {This is the purpose of
  appendix~\ref{sec:calibr-lindbl-torq}, in which we evaluate
  simultaneously the standard torque expressions for the Lindblad
  torque and for the total torque. These expressions are respectively} 
\begin{equation}
\label{eq:10}
 \Gamma^{\rm num,l}_L=-(2.00-0.16\alpha_1+1.11\beta_1)\Gamma_{\rm ref}
\end{equation} 
and
\begin{equation}
 \label{eq:11}
 \Gamma_{\rm tot}^{\rm
   num,l}=-(1.09+0.45\alpha_1+0.53\beta_1)\Gamma_{\rm ref}.
\end{equation} 
where the superscript (which stands for numerical, linear) conveys the method by which the expressions
  have been obtained, that is by means of numerical calculations, as an
  approximation to the linear value, along the lines of what has been
  exposed in section~\ref{sec:relev-numer-simul}.
{We note that Eq.~(\ref{eq:10})} is in good agreement with
Eq.~(14) of \citet{pbck10}.  
It is of interest to confront Eq.~(\ref{eq:10}) to a fully
analytic torque estimate (see appendix~\ref{sec:fully-analyt-torq}
for details), which yields: 
\begin{equation}
  \label{eq:12}
  \Gamma^{\rm linear}_L=-(3.86-0.87\alpha_1+2.09\beta_1)\Gamma_{\rm ref}.
\end{equation}
This estimate differs significantly from Eq.~(\ref{eq:10}), which is
also very close to the estimate obtained by solving numerically the
linear equations \citep{2009arXiv0901.2265P}. Similar discrepancy
between fully analytic results and results obtained by other methods
have already been noticed by \citet{2010ApJ...724..730D}. Still, it
broadly captures some of the salient features of the torque: it is
negative for reasonable values of the surface density or temperature
gradient, a faster decay outwards of the temperature (larger
$\beta_1$) shifts the resonances inwards and renders the torque more
negative (i.e., the coefficient of $\beta_1$ is negative), and for
similar surface density gradients and temperature gradients (which
shift the resonances in almost exactly the same amount), there is, in
the case of a surface density gradient, an additional weighting by the
local surface density which tends to counteract the resonance shift
(i.e., the coefficient of $\alpha_1$ is larger than that of
$\beta_1$). However, whereas Eq.~(\ref{eq:10}) shows that both effects
nearly cancel each other, as the coefficient of $\alpha_1$ almost
vanishes [an effect known as the pressure buffer, shown by
\citet{w97}], it is not quite the case of Eq.~(\ref{eq:12}). A near
cancellation is found, in fully analytic torque estimates, only for
very low values of the smoothing length, and the coefficients of
$\alpha_1$ and $\beta_1$ are found to depend heavily on the smoothing
length \citep[see also][]{mg2004}.  The discrepancy between a fully
analytic estimate and the true linear torque value (given either by
the numerical solution of the linear equations or estimated by means of
adequate numerical simulations) likely arises from the approximation
of the torque cut-off in fully analytical estimates. This cut-off is
given by the Eq.~(\ref{eq:91}), which, together with the expression
for $r_m$, the location of the resonance, determines how the Lindblad
torque behaves for higher order resonances. The denominator of
Eq.~(\ref{eq:91}), in particular, is only approximate \citep{arty93a}.
  
The expression of the corotation torque, given by:
\begin{equation}
\Gamma^{\rm num,l}_C=\Gamma_{\rm tot}^{\rm
  num,l}-\Gamma_{\rm L}^{\rm num,l}
= (0.91-.61\alpha_1+0.58\beta_1)\Gamma_{\rm
  ref}
\label{eq:13}
\end{equation}
must be recast for our needs to:
\begin{equation}
\Gamma^{\rm num,l}_C= (0.61{\cal V}+0.58\beta_1)\Gamma_{\rm
  ref}
\label{eq:14}
\end{equation}
where ${\cal V}=d\log(\Sigma/\omega)/d\log r$ is referred to for
brevity as the vortensity gradient. Eq.~(\ref{eq:13})
and~(\ref{eq:14}) are equivalent in power law disks. However, it is
the form of Eq.~(\ref{eq:14}) that must be used for arbitrary
profiles, as the vortensity gradient ${\cal V}$ can differ
significantly from $3/2-\alpha_1$, as we shall see in
section~\ref{sec:corotation-torque-at}. We shall come back to this
substitution in section~\ref{sec:case-non-vanishing}, where we compare
the effect of higher order derivatives on the torque obtained either
by a fully analytic approach or by numerical simulations. We shall see
that these effects, which affect preferentially lower order
resonances,  are not plagued by the inaccuracies due to the torque
cut-off. This allows a direct comparison of analytics and simulations,
which give essentially the same results, thereby validating Eq.~(\ref{eq:14}).

It is instructive to note that the coefficients of $\alpha_1$ and
$\beta_1$ {in Eq.~(\ref{eq:14})} are {approximately} equal.  One can
notice that Eq.~(11) of \citet{tanaka2002} is suggestive of this
result. Getting rid of all terms of this equation that involve
derivatives in $z$, which do not exist in a two-dimensional disk, one
sees that the pole at corotation for $\eta'_m$ consists of two
contributions: one in the vortensity gradient, and another one that
scales with $2\beta$ (remember that Tanaka et al's $\beta$ is half our
$\beta_1$).  Eq.~(\ref{eq:14}) will be used from now on to subtract
the corotation torque (again considered here as a good approximation
to the linear corotation torque), from the total torque at early times
in calculations where the disk profiles are not power laws, so as to
get the Lindblad torque.

\section{Assumptions and set up}
\label{sec:assumptions-set-up}
\subsection{Wake shift}
\label{sec:wake-shift}
In disks in which the surface density profile or the temperature
profile is not a power law of the radius, the locations of Lindblad
resonances differ from their locations in a power law disk with same
value of $\alpha_1$ and $\beta_1$.  The Lindblad torques, which
sensitively depend on the resonance location, can therefore undergo
large changes, not captured by a standard torque expression.  Our
approach is threefold: we give an analytical order of magnitude of the
effect, we check our predictions by means of dedicated numerical
simulations, and by means of a completely analytic estimate
given by summing the contribution of all Lindblad resonances (see 
appendix \ref{sec:fully-analyt-torq}.) Although this last method yields
results more accurate than our order of magnitude estimate, it does
not provide much insight into the relationship between the local properties
of the density and temperature profiles on one hand, and the Lindblad
torque properties on the other hand. For this reason we find it
relevant to get an estimate of magnitude of the effects of higher
order terms as follows:
we focus on the resonances that most
contribute to the torque, which correspond to azimuthal wave numbers
$m=O(h^{-1})$.  For such values of $m$, we evaluate the resonance
shift, which we improperly call the wake shift.  In order to translate
this wake shift into a torque variation, we firstly calibrate the
one-sided torque dependence on the wake shift by a series of dedicated
calculations, in which we evaluate how the (one-sided) Lindblad torque
changes as a resonance is shifted, by tuning the orbital frequency of
the planet that remains on a circular orbit of fixed radius. This is exposed in
appendix~\ref{sec:calibr-one-sided}. From the standard torque formula
at a given resonance, given at Eq.~(\ref{eq:90}), we see that under a
perturbation of the disk profiles the torque can change either as a
result of the forcing potential (hence a shift, in our case) or as a
result of a change of the ``width'' of the resonance, given by the
denominator in $rdD/dr$. The latter effect is negligible in our case,
and most of the torque change is accounted for by the resonance shift.

We note that our simplified method of evaluating the ``wake shift''
presents an obvious flaw: the major contribution to the Lindblad
torque change may not arise from the resonances that most contribute
to the Lindblad torque. This method gives nevertheless a reasonable
order of magnitude of the torque variation, and, more importantly, it
suggests a functional form for the torque variation due to higher
order derivatives. Once this form has been obtained, one can resort to
numerical simulations to adjust numerical coefficients.

We comment that even though we can obtain the linear Lindblad torque
by a linear analysis, which is self-contained and does not
feature free parameters, the result is not strictly
amenable to a simple form such as the one we are aiming to.  An
obvious reason for that is that the higher order derivatives of the
disk profiles at the location of each Lindblad resonance
feature in the result. The reduction of the Lindblad torque to a
simple approximate expression still involves the identification
of a physically grounded, simple functional form, and the subsequent
adjustment of free parameters. Our final expression should therefore
be regarded as an empirical estimate, even though its functional
dependence is motivated by considerations on the wake shift.

\subsection{Set up and method for numerical checks}
\label{sec:set-up-method}
We undertake in the following two kinds of numerical experiments:
\begin{enumerate}
\item in sections~\ref{sec:case-non-vanishing} and
  \ref{sec:case-non-vanishing-1} we check the dependence of the
  Lindblad torque on a given higher order derivative of the
  temperature (section~\ref{sec:case-non-vanishing}) or surface
  density (section~\ref{sec:case-non-vanishing-1}).
\item In section \ref{sec:torque-behavior-at} we examine the Lindblad
  torque behavior at an opacity transition.
\end{enumerate}
For the first series of numerical experiments, we adopt disk profiles
that are power laws of radius, except on the annulus $a-E<r<a+E$. Over
this annulus, the value of $\beta_i$ or $\alpha_i$ of interest is
constant. The disk profiles outside of the annulus are extrapolated so
that $\alpha_1$ and $\beta_1$ are continuous over the whole
computational domain. We chose $E=2H$ or $E=3H$, in order to cover
the domain over which the exchange of angular momentum between the
disk and the planet takes place.

All numerical simulations were performed with the FARGO
code\footnote{\tt http://fargo.in2p3.fr} \citep{fargo2000,fargo2000b},
in its locally isothermal version.  In all cases we evaluate the
linear total torque by a time average from~$2$ to~$4$ orbital periods,
then we subtract the linear corotation torque expression given by
Eq.~(\ref{eq:14}).  An implicit assumption underlying this procedure
is that the linear corotation torque does not depend upon higher order
derivatives of the temperature or vortensity. This is a reasonable
assumption, as this torque is associated to a singularity at
corotation, whereas the Lindblad torque owes its sensitivity to higher
order derivatives to its non-local character. Note however that the
corotation torque must be cast in terms of the gradient of vortensity
and gradient of temperature. The former, in particular, can bear a
dependency on higher order derivatives of the surface density or
temperature.

The calculations of sections~\ref{sec:case-non-vanishing}
and~\ref{sec:case-non-vanishing-1} were performed with a planet of
mass $M_p=10^{-6}M_*$, embedded in a disk with aspect ratio $h=0.05$
at the planetary orbit. The mesh extends from $R_{\rm min}=0.6$ to
$R_{\rm max}=1.4$, and has resolution $N_{\rm rad}=1200$ and
$N_{\theta}=1500$. Non-reflecting boundary conditions were used, but
those should have no impact on the results, since during the early
stage in which we evaluate the torque the wake has hardly reached the
mesh boundaries.  More details about the setup of
section~\ref{sec:torque-behavior-at} will be given therein.

\section{Individual effects of higher order derivatives}
\label{sec:indiv-effects-high}
\subsection{Expression of the wake shift}
\label{sec:expr-reson-shift}
The ``effective'' location $r_m$ \citep[e.g.][]{arty93a} of a Lindblad
resonance associated to the component of the planetary potential with
$m$-fold symmetry is given by:
\begin{equation}
  \label{eq:15}
  D(r_m)=0,
\end{equation}
where
\begin{equation}
  \label{eq:16}
  D(r) = \kappa^2-m^2(\Omega-\Omega_p)^2+\frac{m^2c_s^2}{r^2}.
\end{equation}
The effect of a perturbation of the sound speed or surface density
profile is to induce a variation $\delta D$ of $D$, which shifts the
resonance location by the amount:
\begin{equation}
  \label{eq:17}
  \delta r_m=-\frac{\delta D}{(\partial D/\partial r)_{r_m}}.
\end{equation}
The rotational equilibrium of the disk reads
\begin{equation}
  \label{eq:18}
  r\Omega^2 = r\Omega_K^2+\frac{\partial_rp}{\Sigma},
\end{equation}
while the expression of the epicyclic frequency is
\begin{equation}
  \label{eq:19}
  \kappa^2=4\Omega\left(\Omega+\frac 12r\frac{d\Omega}{dr}\right).
\end{equation}
Using Eqs.~(\ref{eq:18}) and (\ref{eq:19}) we infer:
\begin{equation}
  \label{eq:20}
  \kappa^2=\Omega_K^2+\frac{4}{r}\frac{\partial_rp}{\Sigma}+r\frac{d}{dr}\left(\frac{\partial_rp}{r\Sigma}\right),
\end{equation}
which can be recast, in a power law disk, as:
\begin{equation}
  \label{eq:21}
  \kappa^2=\Omega_K^2[1+(\alpha_1+\beta_1)(\beta_1-2)h^2].
\end{equation}
Deriving $D$ with respect to $r$, we get:
\begin{equation}
  \label{eq:22}
  \frac{\partial D}{\partial r} = -\frac{3\Omega_K^2}{r}\mp \frac{3\Omega_K^2m}{r}-(2+\beta_1)m^2h^2\frac{\Omega_K^2}{r},
\end{equation}
where henceforth the upper (lower) sign is for the outer (inner) Lindblad
resonance, and where we have retained terms in $m^2h^2$ but not those
in $h^2$. Since the components that most contribute to the Lindblad
torque have $m \sim h^{-1}\gg 1$, we can simplify, for our purpose,
Eq.~(\ref{eq:22}) as follows:
\begin{equation}
  \label{eq:23}
  \frac{\partial D}{\partial r} = \mp \frac{3\Omega_K^2m}{r},
\end{equation}
from which we infer
\begin{equation}
  \label{eq:24}
  \delta r_m=\pm\frac{r\delta D}{3\Omega_K^2m}.
\end{equation}
We contemplate below two distinct cases: (i) the case for which the
rotation profile is altered by a perturbation of temperature, the
surface density remaining constant, and (ii) the opposite case, in
which the profile of temperature remains unchanged, but a perturbation
of surface density is introduced.

\subsection{Case of higher order temperature derivatives}
\label{sec:case-non-vanishing} 
The variation of $D$ under an axisymmetric perturbation of sound speed
(or temperature) is:
\begin{equation}
  \label{eq:25}
  \delta D=\frac{4}{r}\partial_r\delta c_s^2
+r\partial_r\left(\partial_r\delta
    c_s^2\right)-2m^2(\Omega-\Omega_p)\delta\Omega+\frac{m^2}{r^2}\delta
  c_s^2=0,
\end{equation}
where $\delta\Omega$ is the perturbation of angular velocity which can
be related to the pressure perturbation by means of Eq.~(\ref{eq:18}):
\begin{equation}
  \label{eq:26}
  \delta\Omega=\frac{\partial_r\delta c_s^2}{2\Omega r}.
\end{equation}
Using Eq.~(\ref{eq:26}), Eq.~(\ref{eq:25}) can be recast as:
\begin{equation}
  \label{eq:27}
  \delta D=\partial_{r^2}^2\delta
    c_s^2\pm\frac{m\partial_r\delta
    c_s^2}{r}+\frac{m^2}{r^2}\delta c_s^2,
\end{equation}
where we retain only the highest order terms in $m$. The radial shift of
the resonance is thus:
\begin{equation}
  \label{eq:28}
  \delta r_m=\frac {1}{3\Omega_K^2}\left(\pm\frac rm\partial_{r^2}^2\delta
    c_s^2+\partial_r\delta c_s^2\pm \frac mr\delta c_s^2\right)
\end{equation}

We now specialize to the case of a non-vanishing $\beta_2$. Exploiting
the fact that $\beta_2$ is non-negligible whenever $\beta_1$ varies
over a length scale much shorter than $r$, we can transform
Eq.~(\ref{eq:5}) into:
\begin{equation}
  \label{eq:29}
  \beta_2 \simeq \frac{hr^2}{c_s^2}\frac{d^2\delta c_s^2}{dr^2}
\end{equation}
Denoting with $x_m=r_m-r_c$, where $r_c$ is the corotation radius, we
can write:
\begin{equation}
  \label{eq:30}
  \delta r_m=\frac{c_s^2\beta_2}{3\Omega_K^2hr^2}\left(\pm\frac rm
    +x_m\pm\frac mr \frac{x_m^2}{2}\right)
\end{equation}
We denote $m_{\rm max }$ the value of the azimuthal wavenumber for
which the torque is maximal.  Since $m_{\rm max}$ is of order $\sim
2h^{-1}/3$, we substitute the location of the resonance $x_m$ with $\pm
\Delta H$, where we introduce $\Delta \simeq 1$ as a free
parameter. This yields:
\begin{equation}
 \label{eq:31}
 \frac{\delta x_w}{x_w} = \frac{\beta_2h}{3}\left[(u\Delta)^{-1}+1+\frac 12u\Delta\right],
\end{equation} 
where $u=m_{\rm max}h$, and where $x_w=r_w-a$ is the distance of the wake to
the orbit.

We can further transform this expression using the value of $u$ inferred
for instance from Fig.~5 of \citet{2008EAS....29..165M}, which gives
$u\simeq 0.63$. Similarly we take $\Delta=(2/3)(1+u^{-2})^{1/2}\approx
1.25$ \citep{arty93a}. This leads to:
\begin{equation}
 \label{eq:32}
 \frac{\delta x_w}{x_w} \approx 0.89\beta_2h.
\end{equation}

We now turn to the case of a non-vanishing $\beta_3$. We can write:
\begin{equation}
  \label{eq:33}
  \beta_3 \simeq \frac{h^2r^3}{c_s^2}\frac{d^3\delta c_s^2}{dr^3},
\end{equation}
hence for a given resonance we have:
\begin{equation}
  \label{eq:34}
  \delta r_m = \frac{\beta_3}{3r}\left(\pm\frac rm x_m+\frac 12 x_m^2\pm\frac{m}{6r}x_m^3\right),
\end{equation}
from which we infer an approximate expression of the relative shift of
the wake:
\begin{equation}
  \label{eq:35}
  \frac{\delta x_w}{x_w}=\pm\frac{\beta_3h}{3}\left(u^{-1}+\frac
  12\Delta +\frac 16u\Delta^2\right)\approx \pm 0.79 \beta_3h
\end{equation}
We can evaluate the impact of those shifts on the differential
Lindblad torque using the calibration presented in
appendix~\ref{sec:calibr-one-sided}.  We use the approximation that:
\begin{equation}
  \label{eq:36}
  x_{O,I}\frac{\partial\Gamma_{O(I)}}{\partial x_{O(I)}}=-\Lambda\Gamma_{O(I)},
\end{equation}
with $\Lambda\approx 2$, where $\Gamma_{O(I)}$ is the one-sided Lindblad
torque exerted at outer (inner) Lindblad resonances, and where
$x_{O(I)}$ is the distance of the outer (inner) wake to corotation.
We eventually write the outer and inner torques respectively as
\begin{eqnarray}
  \label{eq:37}
 \Gamma_O&=&-\Gamma^0(1+\eta h)\\
  \label{eq:38}
  \Gamma_I&=&\Gamma^0(1-\eta h),
\end{eqnarray}
where $\Gamma^0$ is the one-sided Lindblad torque (average of the
absolute value of the inner and outer torques), and where $\eta$
is a dimensionless parameter that quantifies the degree of asymmetry
between the inner and outer torque. We have the following expression
for the one-sided Lindblad torque:
\begin{equation}
  \label{eq:39}
  \Gamma^0 = 0.4\Sigma\Omega_p^2a^4h^{-3},
\end{equation}
where the numerical coefficient is easily obtained from numerical
simulations. Both two dimensional runs with a softening parameter
$\epsilon=0.6H$ and three dimensional runs with a very small softening
parameter yield the same coefficient of order $0.4$.  We make the
following comments:
\begin{itemize}
\item the coefficient $\eta$ depends on whether the disk considered is
  two or three dimensional, and in the former case it also depends on
  the softening length of the potential. In our set of calculations,
  it turns out to be of order $\eta \simeq 2.2$, but this value is
  highly specific to our case. As we explain below, nevertheless, this
  does not have any impact on the final result.
\item As $\beta_2$ varies the outer and inner wakes both
  recede (or approach) corotation, as can be seen from
  Eq.~(\ref{eq:31}). We therefore expect a minute effect of those shifts
  on the differential Lindblad torque. 
\item On the contrary, terms arising from $\beta_3$ yield shifts of
  outer and inner torques that have same sign, thus giving a much
  stronger (cumulative) effect on the differential Lindblad
  torque. The effect in that case will essentially scale with the
  one-sided Lindblad torque, and the asymmetry term $\eta h$ does not
  feature at lowest order.
\end{itemize}
Using Eq.~(\ref{eq:36}), we have:
\begin{eqnarray}
  \label{eq:40}
 \delta\Gamma_O&=&-\Lambda\Gamma_O\frac{\delta x_O}{x_O}\\
  \label{eq:41}
  \delta\Gamma_I&=&-\Lambda\Gamma_I\frac{\delta x_I}{x_I}
\end{eqnarray}
We consider separately the case of terms of even order, which have
$\delta x_O/x_O=\delta x_I/x_I$, and the case of terms of odd order,
for which $\delta x_O/x_O=-\delta x_I/x_I$. In the first case we find
\begin{equation}
\label{eq:42}
\delta \Gamma_L=-\Lambda\Gamma_L\frac{\delta x_O}{x_O},
\end{equation}
whereas in the second case we get:
\begin{equation}
\label{eq:43}
\delta \Gamma_L=2\Lambda\Gamma^0\frac{\delta x_O}{x_O}.
\end{equation}
In this last case, using Eq.~(\ref{eq:35}), we are left with:
\begin{equation}
  \label{eq:44}
  \delta \Gamma_L\approx 3.17\Gamma_0\beta_3h
\end{equation}
which can be recast as:
\begin{equation}
  \label{eq:45}
  \delta\Gamma_L\approx 1.27 \beta_3\Gamma_{\rm ref}
\end{equation}

\begin{figure}
 \includegraphics[width=\textwidth]{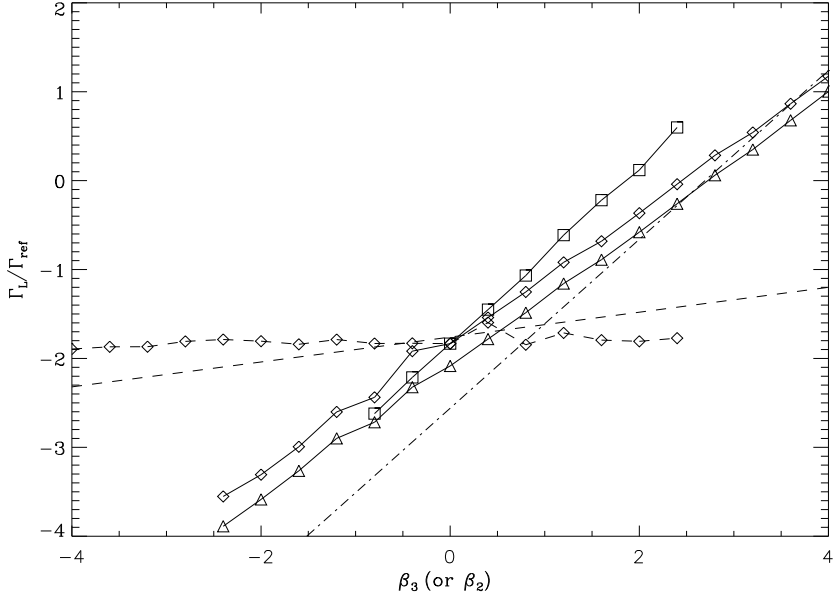}
 \caption{Lindblad torque dependence on the second and third
   derivatives of temperature. Solid lines stand for the case of the
   third derivative ($\beta_3$). The curve with diamonds stands for a
   reference calculation ($\alpha_1=3/2$ and $E=2H$). The curve with
   triangles has $\alpha_1=0$ and $E=2H$, whereas the curve with
   squares corresponds to $\alpha_1=3/2$ and $E=3H$. The dashed lines
   show the dependency of the differential Lindblad torque on the
   second derivative of temperature ($\beta_2$): the line with
   diamonds represents the results of numerical simulations, whereas
   the plain dashed line represents the expectation of
   Eq.~(\ref{eq:48}). Some runs with high values of $|\beta_2|$ or
   $|\beta_3|$, which have extremely high or extremely low sound speed
   near the boundaries, happen to have crashed in the very first time
   steps, which explains why some points are missing in this
   plot. The dot-dashed line shows the linear dependence on
     $\beta_3$ of the fully analytic torque estimate, extrapolated to
     larger values of $\beta_3$, for the case $\alpha_1=3/2$ and
     $E=3H$ (see main text for details.)}
 \label{fig:ld3t}
\end{figure} 
The numerical calculations presented in Fig.~\ref{fig:ld3t} exhibit
a dependency in broad agreement with the estimate of
Eq.~(\ref{eq:45}), as a linear regression fit on this data yields:
\begin{equation}
  \label{eq:46}
  \delta\Gamma_L=K \beta_3\Gamma_{\rm ref},
\end{equation}
where the actual value of $K$, comprised between $\sim 0.75$ and $\sim
1.0$, depends on the details of the setup, and essentially on the
value of $E$ (see Fig.~\ref{fig:ld3t}). This figure also shows the
dependency on $\beta_3$ of the analytic torque estimate (see appendix
\ref{sec:fully-analyt-torq} for details). We note that our fully analytical
estimates do not cover the whole range of $\beta_3$, as in our simple
implementation the algorithm that seeks the location of a resonance
fails for very distorted profiles. Rather, we find that the Lindblad
torque value has a linear dependence on $\beta_3$ for low values of
$\beta_3$ ($|\beta_3| < 0.3$), and that its slope coincides quite
remarkably with that obtained in numerical simulations (we find a
slope $K$ of the analytic estimate of $0.71$ for $E=2H$, and of $0.95$
for $E=3H$). It is surprising, at first, that such a good agreement is
obtained between numerics and analytics, whereas the analytic estimate
performs poorly for power law disk profiles (see
section~\ref{sec:stand-lindbl-torq}). We note, however, that the
further a resonance lies from the orbit, the larger its shift, so that
most of the torque change is due to resonances of order lower than
those which predominantly contribute to the torque. This corroborates
the fact that the inaccuracy of the fully analytic torque expressions
likely comes from highest order resonances, as a consequence of the
approximate form of the torque cut-off. The good agreement that we
find also justifies, indirectly, our substitution from
Eq.~(\ref{eq:13}) to Eq.~(\ref{eq:14}).

It is also noteworthy that this comparison between numerics and
analytics completely fails if one does not include the factor
$\Omega/\kappa$ in Eq.~(\ref{eq:91}). In that case one finds a nearly
flat dependence of the torque on $\beta_3$.

We now work out the torque change due to $\beta_2$. Using
Eqs. (\ref{eq:31}) and (\ref{eq:42}), we are led to:
\begin{equation}
  \label{eq:47}
  \delta\Gamma_L=-\Lambda\Gamma_L\frac{(u\Delta)^{-1}+1+u\Delta/2}{3}\beta_2h\approx -3.37\beta_2h\Gamma_L
\end{equation}
Using the fact that, in our case, we turn out to have $\Gamma_L\sim
-2\Gamma_{\rm ref}$ (the actual coefficient depends on the
$\alpha_i$'s and $\beta_i$'s), we eventually write:
\begin{equation}
  \label{eq:48}
  \Gamma_L\sim 2.7h\beta_2\Gamma_{\rm ref} \sim
  0.14\beta_2\Gamma_{\rm ref}.
\end{equation}
The dependency on $\beta_2$, as expected, is much weaker than the
dependency on $\beta_3$, so we shall neglect it.
Figure~\ref{fig:ld3t} shows that the actual dependence of the Lindblad torque
on $\beta_2$ is even weaker.

\subsection{Higher order derivative of surface density}
\label{sec:case-non-vanishing-1}
We now turn to the dependency of the differential Lindblad torque on
the higher order derivatives of the surface density, assuming that the
temperature profile remains a power law of radius.
For this purpose we recast Eq.~(\ref{eq:20}) as:
\begin{equation}
 \label{eq:49}
 \kappa^2=\Omega_K^2
+(2-\beta_1)\Omega_K^2h^2\frac{\partial\log\Sigma}{\partial\log r}
+\Omega_K^2h^2\frac{\partial^2\log\Sigma}{\partial(\log r)^2}
+\frac 4r\partial_rc_s^2
+r\partial_r\left(\frac{\partial_rc_s^2}{r}\right).
\end{equation}
In a similar fashion, we transform Eq.~(\ref{eq:18}) as:
\begin{equation}
  \label{eq:50}
  r\Omega^2=r\Omega_K^2+r\Omega_K^2h^2\frac{\partial\log\Sigma}{\partial\log
    r}+ \partial_rc_s^2,
\end{equation}
so we can write:
\begin{equation}
  \label{eq:51}
  \delta D = (2-\beta_1\pm m)\Omega_K^2h^2\frac{\partial\delta\log\Sigma}{\partial\log
    r}+\Omega_K^2h^2\frac{\partial^2\delta\log\Sigma}{\partial(\log r)^2}
\end{equation}
Considering as previously that terms that contribute significantly to
the torque have $m \gg 1$, Eq.~(\ref{eq:17}) leads us to:
\begin{equation}
  \label{eq:52}
  \delta r_m=\frac{h^2r}{3}\frac{\partial\delta \log\Sigma}{\partial\log
    r}\pm
\frac{h^2r}{3m}\frac{\partial^2\delta\log\Sigma}{\partial(\log r)^2}
\end{equation}
As in the case of perturbations of the sound speed profile, we
entertain separately the cases $\alpha_2\ne 0$ and $\alpha_3\ne 0$.

In the first case, we have:
\begin{equation}
  \label{eq:53}
  \delta r_m=\frac{h\alpha_2}{3}\left(x\pm\frac rm\right),
\end{equation}
hence
\begin{equation}
  \label{eq:54}
  \frac{\delta
    x_m}{x_m}=\frac{h\alpha_2}{3}\left[1+(u\Delta)^{-1}\right]\approx 0.76h\alpha_2
\end{equation}
This dependency is almost the same as the one predicted by
Eq.~(\ref{eq:31}) on $\beta_2$, and the torque has nearly
same dependence on $\alpha_2$, namely:
\begin{equation}
  \label{eq:55}
  \delta \Gamma_L \approx 0.12\alpha_2\Gamma_{\rm ref}.
\end{equation}
In addition to the effect of the resonance shifts there is an
additional torque variation arising from the change of surface density
itself.  The perturbed surface density corresponding to $\alpha_2$
being even in $x=r-r_c$, its impact on the differential Lindblad
torque is very small and is not contemplated here.

We now evaluate the torque variation due to the third derivative of
surface density.  We firstly work out the resonance shift, and then
evaluate the additional effect arising from the torque weighting by
surface density.

We write:
\begin{equation}
  \label{eq:56}
  \frac{\partial\delta\log\Sigma}{\partial\log r}=\frac 12\frac{x^2\alpha_3}{r_c^2h^2}
\end{equation}
and
\begin{equation}
  \label{eq:57}
  \frac{\partial^2\delta\log\Sigma}{\partial(\log r)^2}=\frac{x}{r_c}\frac{\alpha_3}{h^2},
\end{equation}
hence Eq.~(\ref{eq:52}) gives:
\begin{equation}
  \label{eq:58}
  \delta r_m=\left(\frac{x^2}{6r}\pm\frac{x}{3m}\right)\alpha_3,
\end{equation}
so that:
\begin{equation}
  \label{eq:59}
  \frac{\delta x_m}{x_m}=\pm\frac h3\left(u^{-1}+\frac
    12\Delta\right)\alpha_3\approx 0.73 h\alpha_3
\end{equation}
A calculation similar to that of Eqs.~(\ref{eq:44}) to~(\ref{eq:45})
yields:
\begin{equation}
  \label{eq:60}
  \delta\Gamma_L\approx 1.23\Gamma_{\rm ref}\alpha_3
\end{equation}
In addition to the resonance shift, the change of surface density also
impacts the differential Lindblad torque. One can easily see that the
surface density contribution tends to compensate the wake
shift. Assume that $\alpha_3$ is positive. The resonance shift given
by Eq.~(\ref{eq:59}) shows that the wake is shifted outward, which
yields an increase of the inner torque and an increase of the outer
torque (decrease in absolute value), hence an increase of the
differential Lindblad torque, as seen previously in the case of a
non-vanishing $\beta_3$. However, there is now also a larger surface
density at the outer wake, and a lower surface density at the inner
wake. The contribution of this surface density change is opposite to
that of the resonance shift, hence we expect the dependence of the
differential Lindblad torque on $\alpha_3$ to be weaker than that on
$\beta_3$. We show hereafter that the contribution of the surface
density change is of same order than that due to resonance shift
(albeit with the opposite sign), and then we resort to numerical
calculations to assess the actual dependence.

\begin{figure}
  \includegraphics[width=\textwidth]{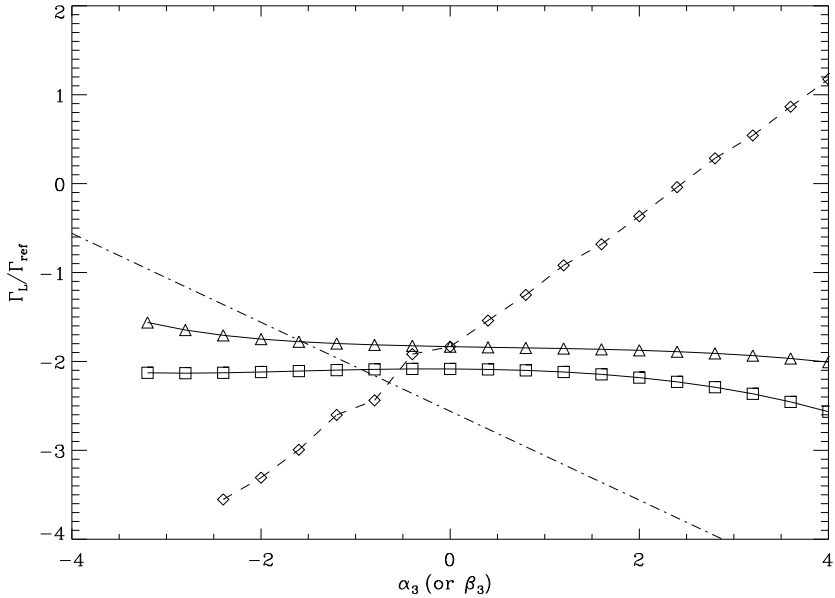}
  \caption{Differential Lindblad torque dependence on $\alpha_3$
    (solid lines) compared to the dependence on $\beta_3$ (dashed
    line, which corresponds to the reference calculation of
    Fig.~\ref{fig:ld3t}).  The differential Lindblad torque turns out
    to be essentially insensitive to $\alpha_3$. Besides, its weak
    dependence on $\alpha_3$ is not a simple linear relationship, and
    its shape depends (slightly) on the value of $\alpha_1$
    ($\alpha_1=0$ for the curve with squares, whereas $\alpha_1=3/2$
    for the curve with triangles). The dot-dashed line shows the
    dependency on $\alpha_3$ of the full analytic estimate of the
    torque, extrapolated from lower values of $\alpha_3$ ($|\alpha_3|
    < 1$, see section~\ref{sec:case-non-vanishing}).}
  \label{fig:dlta3}
\end{figure}

In order to evaluate separately the impact of the surface density
change on the one-sided (then differential) Lindblad, we use the
torque expression of Eq.~(\ref{eq:90}), so that we can write the
relative torque change as:
\begin{equation}
  \label{eq:61}
  \frac{\delta\Gamma^0}{\Gamma^0}=\frac{\sum_m \Gamma_m\delta\Sigma(r_m)/\Sigma(r_m)}{\sum_m \Gamma_m},
\end{equation}
Using Eq.~(\ref{eq:6}), we recast Eq.~(\ref{eq:61}) as:
\begin{equation}
  \label{eq:62}
  \frac{\delta\Gamma^0}{\Gamma^0}=\frac{\alpha_3h}{6}\frac{\sum_m
    \hat x_m^3\Gamma_m}{\sum_m \Gamma_m},
\end{equation}
where
\begin{equation}
  \label{eq:63}
 \hat  x_m=\frac{r_m-a}{H}\approx\pm \frac{2}{3}\sqrt{\frac{1}{m^2h^2}+1}.
\end{equation}
In Eq.~(\ref{eq:62}), whenever a value of $|\hat x_m|$ is
larger than $E/H=3$, the half size of the region over
which $\alpha_3$ is constant and non-vanishing, we clamp it to $\pm
E/H$. This happens for the smallest values of $m$ and is meant to
adhere strictly to the set up of numerical simulations.
The evaluation of Eq.~(\ref{eq:62}) yields
\begin{equation}
  \label{eq:64}
  \frac{\delta\Gamma^0}{\Gamma^0}=1.16\alpha_3h
\end{equation}
We deduce that the surface density impact on the differential Lindblad torque
is, using Eq.~(\ref{eq:39}):
\begin{equation}
  \label{eq:65}
  \delta
  \Gamma_L=\delta\Gamma_O+\delta\Gamma_I=1.16\alpha_3h(\Gamma_O-\Gamma_I)\approx
  -2.32\alpha_3h\Gamma^0\approx -0.93\alpha_3\Gamma_{\rm ref}.
\end{equation}
The net effect of the surface density change, which combines the
effect of the resonance shifts, given by Eq.~(\ref{eq:60}), and the
change of the background density itself, given by Eq.~(\ref{eq:65}),
amounts to:
\begin{equation}
  \label{eq:66}
  \delta\Gamma_L \approx 0.3\alpha_3\Gamma_{\rm ref}
\end{equation}
We see therefore that the effect of the third order derivative of the
surface density should be weaker than that of the third order
derivative of the temperature.  Numerical calculations presented in
Fig.~\ref{fig:dlta3} show that the net dependence is actually much
weaker, which shows that the two effects described respectively in
Eqs.~(\ref{eq:60}) and (\ref{eq:65}) approximately cancel out. As a
consequence, for our purpose we can consider that the differential
Lindblad torque does not depend on $\alpha_3$ (as stressed in section
\ref{sec:wake-shift}, the analytic estimate of the shift of resonances
that most contribute to the torque is used as a guideline and the
actual dependency found in numerical calculations is favored for the
reasons exposed in that section.) Fig.~\ref{fig:dlta3} also shows
  the dependency of the full analytic torque estimate (see
  appendix~\ref{sec:fully-analyt-torq} for details) on $\alpha_3$. The
  dependence has opposite sign that the one inferred from the simple
  calculation based on the wake shift. The surface density weighting
  again overshoots the effect of the resonance shifts. This
  observation is reminiscent of the results for power law profiles
  given in section~\ref{sec:stand-lindbl-torq}, where the coefficient
  of $\alpha_1$ of the fully analytic torque estimate was found to have a
  sign opposite than that of $\beta_1$.

We have seen that the only non-negligible contribution to the
differential Lindblad torque from higher order derivatives of the
density or temperature is the term that stems from the third order
derivative of the temperature. 

We can now propose a tentative generalized differential Lindblad
torque expression in non power-law disks, for the two dimensional
disks with softening length $\epsilon=0.6H$. Out of the four
dependencies that we have worked out (on $\beta_2$, $\beta_3$,
$\alpha_2$ and $\alpha_3$), the dependency on $\beta_3$, given at
Eqs.~(\ref{eq:45})-(\ref{eq:46}), is much larger than the other ones,
given by Eqs.~(\ref{eq:48}), (\ref{eq:55}) and~(\ref{eq:66}). It is
therefore the only contribution that we shall retain to evaluate the
Lindblad torque at locations where the disk profiles are not power
laws, so that our tentative generalized differential Lindblad torque
reads:
\begin{equation}
 \label{eq:67}
 \Gamma_L= -(2.00-0.16\alpha_1+1.11\beta_1-\beta_3)\Gamma_{\rm ref},
\end{equation}
which should be put in contrast with Eq.~(\ref{eq:10}). 

Eq.~(\ref{eq:67}) raises the question of whether one should push the
expansion further so as to include higher order derivatives. This
question can be circumvented in a manner that is exposed in the next
section.

\section{Torque behavior at a temperature transition}
\label{sec:torque-behavior-at}
\subsection{Disk model}
\label{sec:disk-model}
We now consider a disk model in which the surface density is a power
law of radius ($\Sigma\propto r^{-3/2}$), but the sound speed (hence
the temperature) exhibits a more complex radial dependence given by:
\begin{equation}
  \label{eq:68}
  c_s(r) = \frac 12\left[(h_1+h_2)+(h_1-h_2)\tanh\left(\frac{r-r_t}{w}\right)\right]\Omega_K(r)r,
\end{equation}
where $w$, the width of the transition, is given by:
\begin{equation}
  \label{eq:69}
  w=r_th_0
\end{equation}
and where the disk aspect ratios beyond and prior to the transition
are respectively:
\begin{eqnarray}
  \label{eq:70}
  h_1 &=& h_0\times1.1^{1/2}\\
  h_2 &=& h_0\times1.1^{-1/2}\nonumber
\end{eqnarray}
In Eqs.~(\ref{eq:69}) and~(\ref{eq:70}) the fiducial aspect ratio
$h_0$ was chosen to be $h_0=0.04$. The parameters above determine a
sound speed jump of $\sim 10$~\% over a scale length $\sim 2h_0r$,
corresponding to a temperature jump of $\sim 20$~\%.

We note that the dust Rosseland opacity $\kappa_R$ can exhibit
discontinuities of as much as a factor of two as a function of
temperature \citep{1996A&A...311..291H}. Since the mid plane temperature
(which matters for torque calculations) scales with $T_{\rm
  eff}\kappa_R^{1/4}$, where $T_{\rm eff}$ is the effective disk
temperature, which has a smooth profile imposed by the accretion rate
\citep{1994Icar..112..405C}, we can expect jumps of the mid plane
temperature of as much as $2^{1/4} \sim 1.19$. We therefore regard our
simplified disk model as a {\em bona fide} representation of disk
properties at an opacity transition, albeit somehow extreme.

\subsection{Lindblad torque}
\label{sec:lindblad-torque}
\begin{figure}
 \includegraphics[width=\textwidth]{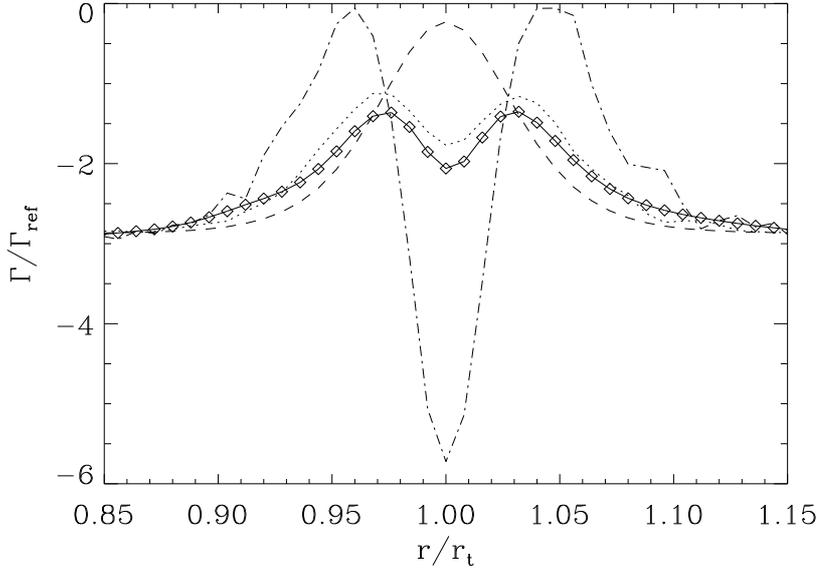}
 \caption{Lindblad torque near the opacity transition, as a function
   of the orbital radius. The solid line shows the data from the
   numerical simulations (each diamond stands for a given run),
   the dashed line shows the results of Eq.~(\ref{eq:10}) and the
   dash-dotted line shows the results of Eq.~(\ref{eq:67}). The
   dotted line shows the result of Eq.~(\ref{eq:79})}
 \label{fig:ot1}
\end{figure}

Figure~\ref{fig:ot1} shows the results of numerical simulations, in
which we sample the transition opacity with 51 different runs. The run
number $i$ ($i\in[0,50]$) considers
 a planet of mass $M_p=10^{-6}M_*$, orbiting on a fixed circular orbit
 of radius $r_i=[0.4\times(i/50)+0.8]r_t$.

We note on this figure that the Lindblad torque displays a negative
peak at the opacity transition, whereas the prediction of the standard
formula for power law disks exhibits a positive peak (as a
straightforward consequence of the decrease of $\beta_1$ at the
transition). Finally, Eq.~(\ref{eq:67}) does reproduce a negative
peak, as expected, but with an exaggerated amplitude.
 
The origin of the discrepancy can be traced back to the conditions
under which Eq.~(\ref{eq:67}) apply. When deriving this equation we
assume that the third derivative of the temperature profile is
constant over the whole region of tidal interaction between the planet
and the disk, that is at least over $[r_p-2H,r_p+2H]$. This is not the
case of the present model, in which the transition occurs on a short
spatial scale of order $\sim H$.

It is therefore more accurate to keep the local values of the first
and second order derivatives at the outer and inner wake, in
Eq.~(\ref{eq:28}), which is therefore recast as:
\begin{equation}
  \label{eq:71}
  \delta r_m\approx\frac{1}{3\Omega_K^2}\left[\pm\frac{c_{s0}^2}{mh}\frac{\beta_2^\pm}{r}-\frac{c_{s0}^2}{r}(\beta_1^\pm-\beta_1^c)\right],
\end{equation}
where $c_{s0}^2$ denotes the sound speed at the orbit, and where the
superscript $\pm$ means that the value must be taken either at the
outer wake ($+$) or at the inner wake ($-$), and where the $c$
superscript means that the value must be taken at the planetary orbit.
We have made use in Eq.~(\ref{eq:71}) of the approximate equalities:
\begin{equation}
  \label{eq:72}
  \partial_r\delta p|^\pm\approx -(\beta_1^\pm-\beta_1^c)\frac pr
\end{equation}

As previously, we leave the exact location at which these values are
to be evaluated as a function of a free dimensionless parameter $D$,
so that variables relative to the outer (inner) wake are to be taken
at $r_c+DH$ ($r_c-DH$).

Using Eq.~(\ref{eq:36}), we have:
\begin{equation}
  \label{eq:73}
  \delta\Gamma_L =\frac{\Lambda\Gamma^0}{H}(\delta x_O+\delta x_I),
\end{equation}
where the shift of the outer wake $\delta x_O$ and that of the inner
wake $\delta x_I$, using Eq.~(\ref{eq:71}), are given respectively by:
\begin{eqnarray}
  \label{eq:74}
 \delta x_O&=&\frac{H^2}{3r}\left[\frac{\beta_2^+}{u}-(\beta_1^+-\beta_1^c)\right]\\
  \label{eq:75}
  \delta x_I&=&-\frac{H^2}{3r}\left[\frac{\beta_2^-}{u}+(\beta_1^--\beta_1^c)\right],
\end{eqnarray}
so that the total variation of the Lindblad torque amounts to:
\begin{equation}
  \label{eq:76}
  \delta\Gamma_L=\frac{\Lambda\Gamma^0h}{3}\left[\frac{\beta_2^+-\beta_2^-}{u}+(2\beta_1^c-\beta_1^+-\beta_1^-)\right].
\end{equation}
We note that both terms in the bracket of Eq.~(\ref{eq:76}) stand for
finite difference approximations of the third derivative of the
temperature, as:
\begin{eqnarray}
  \label{eq:77}
  \beta_2^+-\beta_2^-&\approx& 2D\beta_3\\
\label{eq:78}
2\beta_1^c-\beta_1^+-\beta_1^-&\approx&D^2\beta_3.
\end{eqnarray}
Performing this substitution, we would be led to an equation similar
to Eq.~(\ref{eq:45}). As pointed out above, however, we prefer to keep
the actual values on either side of the orbit, for $\beta_3$ itself
may vary significantly over the region of excitation of the wake. We
note that the coefficient of $\beta_1^c$, not surprisingly, is of the
order of magnitude of the negative of the coefficient of $\beta_1$ in
the standard torque formula. The last term of Eq.~(\ref{eq:76})
therefore amounts to substituting, in the standard torque formula, the
value of $\beta_1^c$ by the average of $\beta_1^+$ and
$\beta_1^-$. This is rather intuitive, as it is the pressure gradient
at these locations that shifts the resonances. In a power law disk,
these three values are equal and the standard formula applies, but if
there is a non vanishing third derivative of the temperature,
$(\beta_1^++\beta_1^-)/2$ may differ from $\beta_1^c$.

This effect, nonetheless, has only the consequence that it smears
slightly the positive peak of Fig.~\ref{fig:ot1}. Most of the effect
is actually due to the first term in the bracket of Eq.~(\ref{eq:76}),
which comes from the second order derivative of the pressure, that
contributes to change the epicyclic frequency and yields most of the
resonance shift. Since the third derivative of the temperature is not
constant over the torqued region, the relative shift of the resonances
with different values of $m$ differs from the case of
section~\ref{sec:case-non-vanishing}, with the consequence that the
coefficient of the dependence on the values of $(\beta_2^+-\beta_2^-)$
may differ, so that we leave it as a free parameter in our final fit.

Incorporating these two changes into Eq.~(\ref{eq:10}), we eventually
obtain the following Lindblad torque expression:
\begin{equation}
  \label{eq:79}
    \Gamma_L=-\left[2.00-0.16\alpha_1+1.11\frac{\beta_1^++\beta_1^-}{2}-0.8(\beta_2^+-\beta_2^-)\right]\Gamma_{\rm ref},
\end{equation}
where we remind that a quantity with a $+$ ($-$) exponent must be
evaluated at $r=a+DH$ ($r=a-DH$), and where $D$ is a free
dimensionless coefficient that we adjusted to get the best agreement
with numerical simulations. We find that we get the best agreement for
$D = 0.2$, and in that case the coefficient of $\beta_2^+-\beta_2^-$
must be $0.8$.  We remind that $\beta_2$ is defined in
Eq.~(\ref{eq:7}).  The results of Eq.~(\ref{eq:79}) are displayed in
Fig.~\ref{fig:ot1}. They show a good agreement with the Lindblad
torque inferred from the numerical simulations. Some comments are in
order about the relevance of the fit given at Eq.~(\ref{eq:79}):
  \begin{itemize}
  \item In the case in which the third order derivative of the
    temperature ($\beta_3$) vanishes over the torqued region, and
    therefore in particular in power law disks, Eq.~(\ref{eq:79})
    reduces to the standard formula of Eq.~(\ref{eq:10}), since in
    that case $\beta_2^+-\beta_2^-=0$, and
    $(\beta_1^++\beta_1^-)/2=\beta_1^c$. Eq.~(\ref{eq:79}) can
    therefore be used everywhere in the disk.
  \item As the planet moves radially and samples different locations
    of the opacity transition, the relative shift of the resonances
    with different values of $m$ varies, so that a relationship such
    as the one proposed can only be approximate. We therefore seek
    agreement where the effect is strongest, that is at the nominal
    position $r=r_t$ of the transition, where the peak of the Lindblad
    torque is observed.
  \item We shall see in section~\ref{sec:total-torque-at} that the
    expression proposed satisfactorily accounts for the torque
    measured for transitions of different widths, and that the
    strength of the effect scales the inverse cube of the
    width. Transitions significantly more narrow than those
    contemplated here are Rayleigh unstable
    \citep{2010MNRAS.402.2436Y}, whereas the effect will be negligible
    in transitions significantly wider. Eq.~(\ref{eq:79}) constitutes
    therefore an acceptable description of the effect for all
    situations in which it is measurable. 
  \item Contrary to what has been done with the temperature slopes, the
    surface density slope coefficient is left unchanged. There are
    several reasons for this: (i) the coefficient of $\alpha_1$ is
    small, hence no measurable change is expected if one replaces its
    value with the average of its values at the outer and inner wake;
    (ii) we have seen in section~\ref{sec:case-non-vanishing-1} that the
    additional effect of surface density weighting of the torque tends
    to cancel out the effect of the resonance shifts, hence we do not
    write a term in $\alpha_2^+-\alpha_2^-$, similar to that in
    $\beta_2^\pm$, since that term would have a small coefficient, and
    (iii) at an opacity transition, no significant effect is expected for
    the surface density, counter to the temperature.
  \item The changes with respect to the standard formula consists in:
    (i) an evaluation of the shift of the resonances using the slope at the
    wake location itself, rather than at the orbit (this
    corresponds to the term in $\beta_1^++\beta_1^-$ and it is a small
    effect), and (ii) an evaluation of the change of the epicyclic frequency
    itself, the value of which defines the location of Lindblad
    resonances (this corresponds to the term in $\beta_2^+-\beta_2^-$,
    and it is the most important effect). From Eq.~(\ref{eq:20}), we
    see that the derivative of the temperature to second order suffices
    to evaluate the change in the epicyclic frequency, and no further
    expansion is needed, provided the derivatives are evaluated at a location
    intermediate between the orbit and the wake, rather than at the orbit.
  \end{itemize}
  
\subsection{Comparison to the fully analytic torque expression}
\begin{figure} \includegraphics[width=\textwidth]{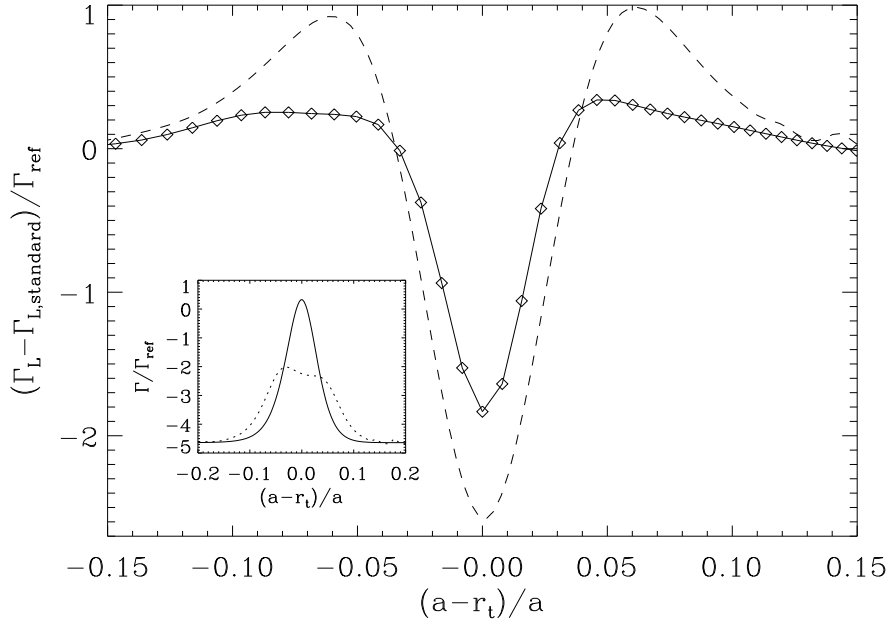} 
\caption{Difference
      between the true value of the torque and the one inferred from
      the standard torque formula, as a function of the distance to
      the nominal radius of the transition. The solid line represents
      the results for numerical simulations (each diamond represents a
      run), and the dashed line shows the results for the analytical
      estimate. The inset plot shows the total Lindblad torque as a
      function of the distance of the nominal radius of the
      transition, obtained using a fully analytic estimate (dotted
      curve) or using the corresponding standard torque expression
      given by Eq.~(\ref{eq:12}).  We recover the behavior found from
      numerical simulations at Fig.~\ref{fig:ot1}, namely that the
      actual profile is wider, and displays a reversed peak at the
      top. The dashed line of the main plot is essentially the
      difference between these two curves.}
  \label{fig:compcav}
\end{figure}
We show in Fig.~\ref{fig:compcav} the difference between the actual
torque value and the value inferred from standard torque
estimates, both numerical simulations and for fully analytic
estimates. Differently said, this differences assesses the magnitude
of the effect of higher order terms of the temperature
profile. Although the global shape of the curve of the fully analytic
expression is broadly similar to that obtained from simulations, it
displays a larger amplitude, which is consistent with the fact that
the standard torque expression obtained by summing torque estimates on
all resonances, given by Eq.~(\ref{eq:12}), has a larger coefficient
for $\beta_1$ than the standard torque expression given by numerical
calculations, given at Eq.~(\ref{eq:10}).  We had noted in
section~\ref{sec:case-non-vanishing} the excellent agreement between
the linear estimate and numerical simulations for the dependence of
the torque on $\beta_3$. Our tentative explanation was that the effect
was primarily due to relatively low order resonances, which had larger
shifts. Here, the resonances that are most shifted are the ones which
lie the closest to the orbit, {\em i.e.} those of higher order.

\subsection{Corotation torque at the transition}
\label{sec:corotation-torque-at}
We now contemplate the other torque component, namely the corotation
torque, which has been thus far omitted.  We examine here the
vortensity related corotation torque, in the linear limit. We
therefore focus on the indirect effects of the temperature profile: it
alters the rotation profile and therefore the vorticity profile, hence
the vortensity gradient.

We wish to evaluate the dimensionless coefficient
\begin{equation}
  \label{eq:80}
  {\cal V} = \frac{d\log(\Sigma/\omega)}{d\log r}.
\end{equation}
Since the profile of $\Sigma$ is a power law, we focus on the part
related to the vorticity $\omega$, given by Eq.~(\ref{eq:2}).
We use Eq.~(\ref{eq:18}), and we keep only the derivatives of
the sound speed, which varies over the length scale $H$, whereas
other quantities vary over $r$.
This yields:
\begin{equation}
  \label{eq:81}
  \partial_r\omega \approx -\frac{\partial_{r^3}^3c_s^2}{2\Omega}.
\end{equation}
Considering that $\omega\sim \Omega/2$ everywhere in the disk,
we are led to:
\begin{equation}
  \label{eq:82}
  {\cal V} \sim -\beta_3.
\end{equation}
Eq.~(\ref{eq:82}) is an approximate relationship that holds wherever
there is a significant peak of $\beta_3$ in the disk. In the power law
parts of the disk, we have ${\cal V}=3/2-\alpha_1$.  We can deduce the
corotation torque at the transition using Eqs.~(\ref{eq:14})
and~(\ref{eq:82}).  This gives:
\begin{equation}
  \label{eq:83}
  \Gamma_C \sim 0.6(\beta_1-\beta_3)\Gamma_{\rm ref}
\end{equation}
Notwithstanding its standard dependency in $\beta_1$, the corotation
torque should therefore display a boost opposite to that of the
Lindblad torque at an opacity transition, as we can see by comparing
Eqs.~(\ref{eq:83}) and~(\ref{eq:46}). Although it may seem at first
glance that the balance is in favor of the Lindblad torque, such is
not the case for the following reasons:
\begin{itemize}
\item the previous section has shown that the peak of Lindblad torque
  observed at the transition was less than that predicted by
  Eq.~(\ref{eq:46}). This was understood as being due to the fact that
  the third derivative is not uniform over the torqued region.
\item In contrast, it is the third derivative of the temperature {\em
    at the orbit} that gives the corotation torque.
\item In addition to this, we remind that the torque value in these
  two dimensional calculations is very smoothing dependent. The value
  of the smoothing adopted here ($0.6H$) is adapted to the Lindblad
  torque, in the sense that it provides a one-sided Lindblad torque of
  the order of that in a three-dimensional disk
  \citep{1999ApJ...516..451M,masset02}. However, this value of the
  smoothing tends to underestimate the corotation torque, for which a
  smoothing length of $\sim 0.4H$ should rather be used (Casoli \&
  Masset, in prep.)
\item Finally, the corotation torque, depending on the planet mass, on
  the disk thickness and on the amount of viscous diffusion,
  eventually remains linear at all times, or settles into the
  so-called horseshoe regime. The horseshoe drag bears same
  dependencies on the disk parameters as the linear corotation torque,
  but it has a larger value \citep{2009arXiv0901.2265P}. Only for very
  small values of the disk's effective viscosity does the horseshoe
  drag enter a so-called saturated regime, so that the total torque
  essentially amounts to the differential Lindblad torque
  \citep{2010ApJ...723.1393M}.
\end{itemize}

\subsection{Total torque at the transition}
\label{sec:total-torque-at}
\begin{figure}
 \includegraphics[width=\textwidth]{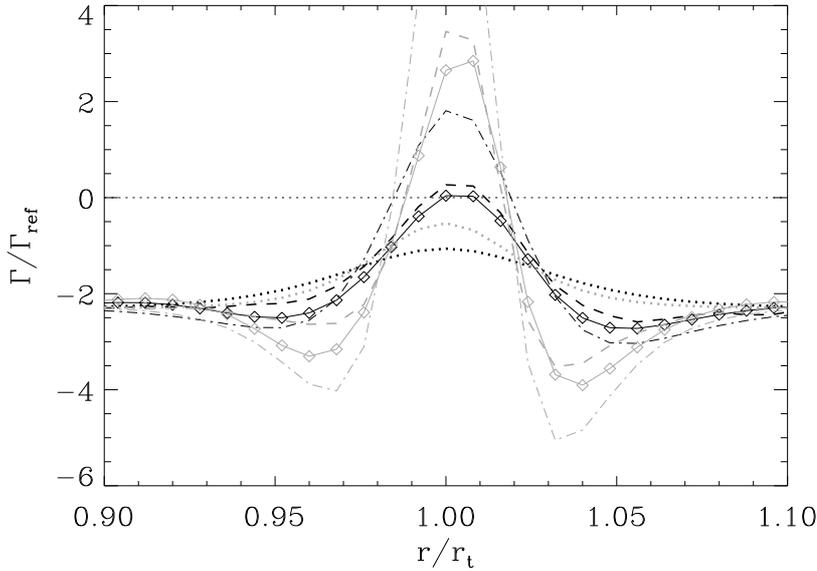}
\caption{Total torque near the opacity transition, as a function of
   the orbital radius. Black lines show the results for the standard
   transition whereas grey lines show the results for the narrow
   transition. See the main text for the signification of each line
   style.}
 \label{fig:totq14}
\end{figure}

We present the total torque at the opacity transition in
Fig.~\ref{fig:totq14}. Results are presented for two transitions: the
same as the one contemplated in previous paragraphs, defined by
Eqs.~(\ref{eq:68}) to~(\ref{eq:70}), which has a width $r_th_0\sim H$,
and a more narrow one ($w_t=0.7r_th_0\sim 0.7H$), which is
nevertheless still largely Rayleigh stable \citep{2010MNRAS.402.2436Y}.
The solid curves with diamonds show the results directly obtained from
numerical simulations (each diamond stands for a different run). The
dotted lines show the expectations of the standard total linear torque
expression of Eq.~(\ref{eq:11}), and the dot-dashed lines show the sum
of Eqs.~(\ref{eq:10}) and (\ref{eq:14}), that is the standard total
torque expression, except that we use the proper value of ${\cal V}$
instead of $3/2-\alpha_1$. Finally, the dashed lines show the sum of
Eqs.~(\ref{eq:14}) and~(\ref{eq:79}).

We see that the standard torque formula fails at reproducing the
strong peak of the total torque. Even in the narrow case, it does not
predict the observed torque reversal. On the contrary, the mere
addition of Eqs.~(\ref{eq:10}) and (\ref{eq:14}) overshoots the real
torque, essentially because it disregards the Lindblad torque
sensitivity on higher order derivatives of the temperature.  Finally,
the dashed lines, that corresponds to the sum of the proper expressions
of the Lindblad and corotation torque, are in excellent agreement with
the torque measurements. They correctly predict the occurrence of
torque reversal, and the magnitude of the peak at the transition.

We also note the steep dependence of the peak on the transition width
$w_t$.  The peak value scales indeed with $\beta_3$, which is
proportional to $w_t^{-3}$.

\section{Discussion}
\label{sec:discussion}
\subsection{Extension of results to more general cases}
We have seen that both the Lindblad torque and the corotation torque
have a dependency on the third derivative of the temperature, with
opposite signs. The balance is in favor of the corotation torque,
which, even in the linear regime, could be enough to stall migration
at opacity transition, provided the transition is sufficiently narrow (we
comment that our calculations, which exhibit such stalling, are
conservative in the sense that they underestimate the corotation
torque by the use of a large softening of the potential).  Early work
on this kind of traps \citep[see][at section 5]{trap06} did not
evidence this possibility of stalling, essentially because the
transition was too large ($\sim 4H$) and the potential softening value
even larger ($\epsilon = 0.7H$).

In the general case, one should take proper care of the value of the
corotation torque by incorporating possible non-linear effects, namely
the horseshoe drag regime and its partially saturated state. Details
about this kind of procedure can be found in
\citet{2010ApJ...723.1393M}.

A result similar to the one presented above is expected to hold for
three dimensional disks. Namely, the two effects exhibited in
section~\ref{sec:torque-behavior-at} should be introduced in the same
manner as for two dimensional disks:
\begin{itemize}
\item The temperature gradient at the orbit should be replaced by its
  average at the outer and inner wakes.
\item A term in $\beta_2^+-\beta_2^-$ should be added. We remarked in
  section~\ref{sec:case-non-vanishing} that the one sided Lindblad
  torque has same value for three dimensional disks and for the two
  dimensional disks that we considered. Since the effect arising from
  the change of epicyclic frequency is a cumulative effect (it has
  same sign for the inner and outer torques) that scales with the
  one-sided Lindblad torque, we expect the coefficient of this
  term to be the same as in our two dimensional disks, that is to say
    $\sim 0.8$.
\end{itemize}
Still, a standard Lindblad torque formula in three dimensional disks
has not been worked out yet. The work of \citet{tanaka2002} is limited
exclusively to globally isothermal disks. The calibration exposed in
appendix~\ref{sec:calibr-lindbl-torq} is hardly achievable in three
dimensional calculations. In particular, one cannot resort to nested
meshes centered on the planet, as the low resolution in the parts of
the horseshoe region that lie far from the planet would act as a
source of diffusion, which would impede a full saturation of the
corotation torque. The tentative Lindblad torque formula given by
\citet{2010ApJ...723.1393M} at Eq.~(160) was based upon three
dimensional calculations in the linear regime for which the total
torque exhibited a dependency in $0.4\beta_1$, and the same dependency
was assumed for the Lindblad torque by ignoring a possible dependency
of the linear corotation torque on $\beta_1$. The present work
stresses that the linear corotation torque has a dependency far from
negligible on the temperature gradient. Work is in progress to assess
this dependence in three dimensional disks (Casoli \& Masset, in
prep.). It is likely that the Lindblad torque has also in that case a
steep dependence on $\beta_1$ \citep{pbck10}.

We stress that our additional term is given in order of magnitude
only, but it is straightforward to apply, and it yields a value quite
close to that obtained from numerics in the regions of the disk where
the profiles cannot be approximated by power laws.

\subsection{Comparison with the analysis of \citet{mg2004}}

In their analysis of the migration of low-mass protoplanets near
opacity transitions, \citet{mg2004} found a very strong slowing down
of migration, corresponding almost to a cancellation of the Lindblad
torque at some locations in the disk. On the contrary, we find here a
relatively mild effect on the Lindblad torque, even though we have
considered a rather extreme opacity transition, both in terms of
narrowness and temperature jump, and even though we consider a smaller
softening length of the potential than \citet{mg2004}. In order to
understand the reason for this discrepancy, we have tried to reproduce
the results of \citet{mg2004} for power law profiles and with a
softening length of the potential equal to $H$. These results were
subsequently recast by \citet{2010ApJ...724..730D}, in their Eq.~(8),
as:
\begin{equation}
  \label{eq:84}
  \Gamma_L^{(\epsilon=H)}=-(0.80-0.77\alpha_1+1.12\beta_1)\Gamma_{\rm ref}.
\end{equation}
The most striking feature of this expression is the small value of
its constant coefficient ($0.80$). This makes the torque value prone
to cancellation even for moderate values of $\alpha_1$ or
$\beta_1$. Using our analytic approach detailed in
appendix~\ref{sec:fully-analyt-torq}, we obtain
\begin{equation}
  \label{eq:85}
  \Gamma_{L,1}^{(\epsilon=H)}=-(2.53+0.32\alpha_1+0.83\beta_1)\Gamma_{\rm
 ref},
\end{equation}
which differs substantially from Eq.~(\ref{eq:84}). There are some
differences in the approach of \citet{mg2004} which may account for
this difference:
\begin{itemize}
\item They use a softening length equal to the disk thickness at the
  location of the resonance, rather than at the location of the planet.
\item They use the adiabatic sound speed in all terms related to
  non-axisymmetric disturbances (it is however unclear
  whether they do use it: their figure~2 shows a virtually perfect
  agreement with the results of \citet{w97}, when they normalize the
  torque to the same quantity, related to the isothermal sound speed
  --~their equations~(15) and~(20)~--; one would expect the torque
  value to be reduced by a factor of $\gamma$ ---~the ratio of the
  specific heats~--- if the adiabatic sound
  speed were used).
\end{itemize}
If we incorporate these changes into our analytic approach, we are
left with the following torque expression:
\begin{equation}
  \label{eq:86}
  \Gamma_{L,2}^{(\epsilon=H)}=-(1.86+0.24\alpha_1+0.83\beta_1)\Gamma_{\rm
 ref}.
\end{equation}
We have tried other combinations (e.g., using a radially varying
smoothing length and using the isothermal sound speed), but none of
them has yielded results close to Eq.~(\ref{eq:84}). In all cases the
constant term remains of order $\sim 2$. The reasons for this
discrepancy are unknown. Their approach differs in that they use a
continuous torque density, which may have a lower accuracy for the
relatively thick disk ($h=0.07$) that they consider, and for the large
value of the potential softening length, as only relatively low order
resonances contribute to the torque in that case. In the same vein,
the transformation of the Laplace coefficients to modified Bessel
functions of the second kind may further degrade the accuracy, but
none of these explanation sounds plausible at accounting for the large
difference between Eq.~(\ref{eq:84}) and Eq.~(\ref{eq:86}). We believe
that the huge effects that they get at opacity transitions are linked
to the smallness of the constant coefficient in Eq.~(\ref{eq:84}).
Our results, both from numerical simulations and from analytic
estimates, are quite different: we find that the effects of opacity
transitions (even extreme ones) on the Lindblad torque cannot reduce
it by more than a factor of about $\sim 2$. However, the total torque
can still be reverted at these locations, not by the Lindblad torque
but by the corotation torque that can undergo a boost originating from
the strong departure from Keplerianity of the flow at the transition.

\section{Conclusions}
\label{sec:conclusions}
We have obtained an expression for Lindblad torques, given at
Eq.~(\ref{eq:79}), that generalizes the standard expression of
Eq.~(\ref{eq:10}) wherever the disk profiles are not power laws, in
particular at opacity transitions. Essentially one additional term
plays a role, at locations where the third radial derivative of
temperature does not vanish. The main effect comes from the alteration
of the epicyclic frequency, which depends on the flow's vorticity. The
latter differs significantly from its nearly Keplerian value wherever
the pressure gradient varies over a short length scale. The resulting
variation of the epicyclic frequency shifts the planetary wake and
yields a substantial change of the Lindblad torque.

Our expression has been obtained for locally isothermal disks. In
disks for which an adiabatic expression is better suited for the
Lindblad torque, the value found here can simply be divided by
$\gamma$ \citep{bm08,pp08,2010ApJ...723.1393M,pbck10}, whereas the
value of $D$ should be multiplied by $\gamma^{1/2}$. The transition
from the isothermal value to the smaller adiabatic value, as a
function of the thermal diffusivity, is described by
\citet{2010ApJ...723.1393M} at Eqs.~(154)-(156), by
\citet{2010ApJ...715L..68L} at Eq.~(10), or by \citet{pbk11} at
Eqs.~(45)-(47). These three procedures are in broad agreement.

There are side results to this analysis. Firstly, we find evidence
{from numerical simulations} that the linear corotation torque in
a locally isothermal disk includes a term that scales with the
temperature gradient, and that the dependence is as steep as the
dependence on the vortensity gradient, which is an important fact {\em
  per se} but which has received little attention so far. Secondly, we
find that the corotation torque, explicitly neglected in the analysis
of \citet{mg2004}, displays a boost at an opacity transition, because
there the vortensity gradient significantly differs from its Keplerian
value $3/2-\alpha_1$. This boost may be sufficient to halt migration
at opacity transitions. An extremum of vortensity is nonetheless
unstable to a Rossby wave instability \citep{lovelace99,li2000}, and
viscous diffusion, either laminar or turbulent, acts at spreading
radially narrow vortensity features. As this was not our primary goal,
these effects and their interplay with migration have been disregarded
in this work.

The present analysis also disregards any adiabatic effects on the
corotation torque. Should the flow behave nearly adiabatically on the
time scale of horse shoe U-turns, the additional entropy related
corotation torque should be taken into account.

\begin{acknowledgements}
  I thank Steven Lubow for pointing me out that the linear corotation
  torque could depend on the temperature gradient.
\end{acknowledgements}

\appendix

\section{Calibration of the Lindblad and linear corotation torques in
  two-dimensional disks with a softened potential}
\label{sec:calibr-lindbl-torq}
{For the Lindblad torque calibration we perform}
long term inviscid calculations
in which we let the horseshoe drag saturate, so that the asymptotic
torque value is considered as a good approximation to the {linear}
Lindblad torque {(see section~\ref{sec:relev-numer-simul}).}
In order to obtain the coefficients $k_0$ to $k_2$ of Eq.~(\ref{eq:8})
for our disks, {and similar coefficient $k_0^t$ to $k_2^t$ for the
total torque} we have run four different calculations, in power law
disks. Three calculations should be sufficient, but we have run a
fourth one as a control to get some confidence that the Lindblad
torque can indeed take the form given by Eq.~(\ref{eq:8}).

Our calculations were carried out for a disk with $h=0.04$ and a
planetary mass $M_p=10^{-6}M_*$. The mesh extends from $R_{\rm
  min}=0.5a$ to $R_{\rm max }=1.6a$, while its resolution is $N_{\rm
  rad}=1650$ and $N_\theta=1000$. This resolution is meant to give the
same cell size as in the calculations of
section~\ref{sec:indiv-effects-high}, but with mesh boundaries which
lie further from the orbit. The horseshoe zone is therefore covered by
approximately $9$ cells radially. The dimensionless parameter that
controls the non-linearity of the flow is $M_p/(h^3M_*) \approx
0.015$.

\begin{figure}
 \includegraphics[width=\textwidth]{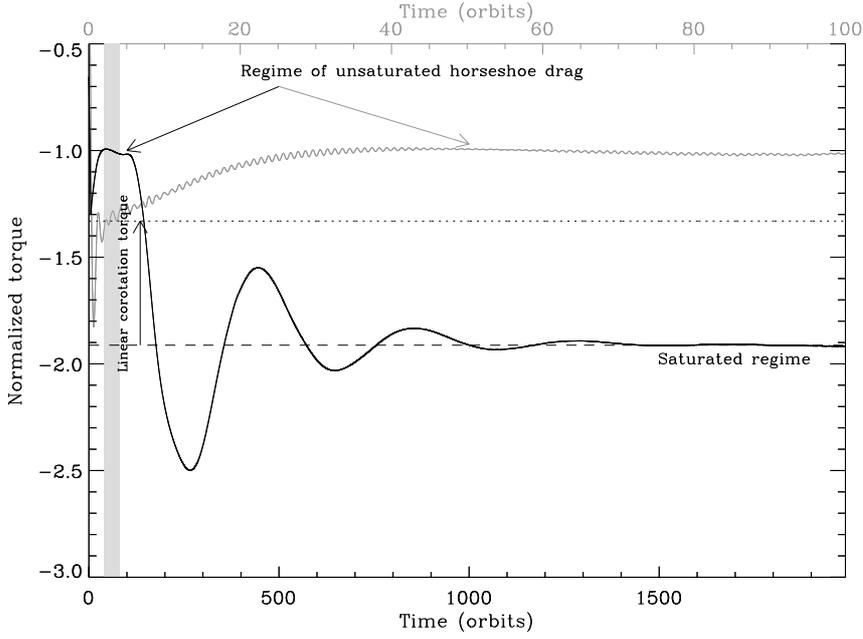}
 \caption{Total torque as a function of time. The torque is shown with
   two different scales as a function of time: over a long time scale
   to exhibit its asymptotic value represented by a dashed line (black
   curve, lower $x$-axis), and over a short time scale to expose the
   transition \citep[see][]{2009arXiv0901.2265P} from the linear
   corotation torque to the non-linear, horseshoe drag regime (grey
   curve, upper $x$-axis). The light grey stripe on the left
   corresponds to the temporal window $2-4$~orbits, over which the
   torque is averaged, and which yields an estimate of the Lindblad
   torque plus linear corotation torque, represented by the dotted
   line.}
 \label{fig:lctcalib}
\end{figure}

In addition, a time average from~$t=2$ to~$4$ orbital periods only of
the total torque gives an estimate of the linear torque value in each
case. We comment that over this time frame the torque can still be
considered as linear. {Indeed, the characteristic timescale for
  the onset of non-linear effects is the horseshoe libration time,
  multiplied by twice the disk aspect ratio
  \citep{2009arXiv0901.2265P}, which can be cast as:
  \begin{equation}
    \label{eq:87}
    \tau_{nl}\sim \frac{8\pi}{3\Omega}(q/h^3)^{-1/2},
  \end{equation}
a timescale which can be identified with the time it takes to execute
a horseshoe U-turn \citep{bm08}. \citet{2009arXiv0901.2265P} mention
that the linear results are valid at very early time (about two
orbits) for their fiducial calculation with $q/h^3=0.1$. For our value
of $q/h^3$, it should be valid over approximately up to $5$~orbits,} 
while the {establishment} of the horse shoe drag
regime takes about $30$~orbits in our case.  The results of one of the
runs are represented in Fig.~\ref{fig:lctcalib}, in which one sees the
three stages of the torque : (i) linear stage, at very early times, up
to a few orbits, (ii) unsaturated horseshoe drag, reached after around
30 orbits, and maintained up to about 120 orbits, and (iii) the
saturation phase, in which the torque oscillates about the
linear Lindblad torque, which constitutes its asymptotic value at
time larger than $10^3$ orbits.

The torque values, normalized to the reference value of
Eq.~(\ref{eq:9}), are presented in Tab.~\ref{tab:sdkfj}.
\begin{table}
\begin{center}
\begin{tabular}{|c|c|c|c|}
\hline
   $\alpha_1$ & $\beta_1$ & $\Gamma_L/\Gamma_{\rm ref}$ & $\Gamma_{\rm
   tot}^{\rm linear}/\Gamma_{\rm ref}$ \\ \hline
   $3/2$ & $0$ & $-1.74$ & $-1.78$ \\ \hline
   $1/2$ & $0$ & $-1.91$ & $-1.32$ \\ \hline
   $3/2$ & $1$ & $-2.86$ & $-2.30$ \\ \hline
   $1/2$ & $1$ & $-3.03$ & $-1.84$ \\ \hline
\end{tabular}
\end{center}
\caption{Normalized Lindblad torque (at larger time) and total linear
  torque (averaged between $2$ and $4$~orbital periods) for the four
  calibration runs.\label{tab:sdkfj}}
\end{table}
{This data yields an over-constrained system of six unknowns
  ($\gamma_0$, $\gamma_1$, $\gamma_2$, $\gamma_0^t$,
  $\gamma_1^t$, $\gamma_2^t$) and eight relations. We add a
  constraint, namely $k_0=(3/2)(k_1^t-k_1)+k_0^t$, to ensure that the
  corotation torque scales exactly with $(3/2)-\alpha_1$ in power law
  disks (while we leave its dependence on the temperature gradient
  as a free parameter), 
  as we need to recast it in terms of the vortensity gradient, which
  differs substantially from $(3/2)-\alpha_1$ when higher order
  derivatives of the surface density or temperature do not vanish. We
  note that the linear corotation torque, obtained from the
  semi-analytic work of \citet{tanaka2002}, scales indeed exactly as
  $(3/2)-\alpha_1$ in power law, two dimensional
  disks\footnote{Although this dependence is not given explicitly for
    two-dimensional disks by \citet{tanaka2002}, one can infer it by
    subtracting the Lindblad torque from the total torque.}.}
  From this data we infer the linear regressions given in
  Eqs.~(\ref{eq:10}) {and} (\ref{eq:11}). We note that if
  we do not impose the constraint given above, we are led respectively
to:
\begin{equation}
\label{eq:88}
 \Gamma_L=-(2.00-0.16\alpha_1+1.12\beta_1)\Gamma_{\rm ref}
\end{equation} 
and
\begin{equation}
  \label{eq:89}
  \Gamma_{\rm tot}=-(1.09+0.46\alpha_1+0.52\beta_1)\Gamma_{\rm ref}
\end{equation}
in stead of Eqs.~(\ref{eq:10}) and~(\ref{eq:11}). The comparison gives
an idea of the fitting error, as does the first line of
Tab.~\ref{tab:sdkfj}. Since there is no vortensity gradient
  and the disk is globally isothermal, both torque values in
  this row should coincide. Sources of inaccuracies may include:
\begin{enumerate}
\item the narrowness of the time frame over which we average the
  torque to get the total linear torque (but taking a larger time
  frame would raise the issue of the corotation torque beginning to be
  non-linear in the presence of a vortensity or temperature gradient);
\item numerical diffusion of the vortensity within the horseshoe
  region, which may act as a physical viscosity and may leave a tiny
  corotation torque at larger time (almost fully saturated, but not
  quite);
\item Partial reflection of the wake on the mesh boundaries, which is
  present at larger time, but not at earlier time when the linear
  total torque is evaluated. As non-reflecting boundary conditions are
  implemented in FARGO, the amount of reflection should be small.
\end{enumerate}

\section{Fully analytic torque estimate}
\label{sec:fully-analyt-torq}
Our method for estimating the Lindblad torque for each azimuthal
component of the planetary forcing potential is essentially based upon
that of \citet{w97}, with two amendments described below.
The Lindblad torque at a Lindblad resonance of $m$th order is given by
\begin{equation}
  \label{eq:90}
  \Gamma_m=\frac{\pi^2m\Sigma(r_m)\Psi_m^2}{r_m(dD/dr)_{r_m}},
\end{equation}
where $r_m$ and $D(r)$ are defined respectively at Eqs.~(\ref{eq:15})
and~(\ref{eq:16}), and where
\begin{equation}
  \label{eq:91}
  \Psi_m=\frac{rd\phi_m/dr+2(\Omega/\kappa)mf\phi_m}{\sqrt{1+4\xi^2}},
\end{equation}
with $\xi=mc_s(r)/r\kappa(r)$, where the epicyclic frequency $\kappa$
is defined at Eq.~(\ref{eq:19}), and where
$f=m[\Omega(r)-\Omega_p]/\Omega(r)$. Unlike \citet{w97}, and
following \citet{mg2004}, we have incorporated to Eq.~(\ref{eq:91})
the ratio $\Omega/\kappa$. The latter turns out to be unimportant in
disks with power law profiles, but it is crucial to include it in
disks with abrupt changes, where the epicyclic frequency may
significantly differ from its nearly Keplerian value. Our second modification
with respect to Ward's formulation is that we include the potential
softening length in the evaluation of the amplitude of the $m$th order 
Fourier component of the disturbing function:
\begin{equation}
  \label{eq:92}
  \phi_m=-\frac{GM_p}{a}b_{1/2,(\epsilon/a)}^m(r/a),
\end{equation}
where
\begin{equation}
  \label{eq:93}
  b_{1/2,(\epsilon/a)}^m(r/a)=\frac{2}{\pi}\int_0^\pi\frac{\cos m\theta d\theta}{\sqrt{1-2(r/a)\cos\theta+(r/a)^2+(\epsilon/a)^2}}.
\end{equation}
The form of Eq.~(\ref{eq:91}), together with the definition of $r_m$
and the definition of the forcing function of Eq.~(\ref{eq:93}),
determines how the torque behaves in the limit of high order
resonances, and therefore incorporates the torque cut-off.  In the
particular case of power law disks, using our fiducial smoothing
length $\epsilon=0.6H$, we find $\Gamma_m\propto \exp(-1.8hm)$ for
larger $m$.  In Eq.~(\ref{eq:93}), the value of the softening length
$\epsilon$ is fixed, and it is thus the same for all Lindblad
resonances. This is consistent with the formulation of
\citet{2009arXiv0901.2265P} and \citet{pbck10} (see in particular
Eq.~(3) of the former). This also corresponds to the standard
implementation of the FARGO code, in which the material in each zone
of the computational domain is acted upon by the planetary potential,
which has a given smoothing length, and reacts back on the latter
through a potential which has the same smoothing length, so as to
approximately ensure the fulfillment of the action and reaction
law. This is in contrast with the formulation of \citet{mg2004}, who
adopt a smoothing length that vary with $r$ (more precisely their
smoothing length is the vertical scale height of the disk at the
location of the resonance).

We further comment that we do not reduce the expression of the Laplace
coefficients (modified by the presence of the smoothing) to
expressions involving modified Bessel functions of the second kind, as
is a standard practice. We rather evaluate directly the integral of
Eq.~(\ref{eq:93}) at each resonance, as well as its radial derivative.
Also, we do not use the convenient device consisting in working out a
torque density, which converts a discrete problem into a continuous
one \citep[e.g.][]{w97}. Rather, we evaluate $dD(r)/dr$ at each
resonance, the location of which is formerly determined using a
Newton-Ralphson method.

\begin{figure}
  \includegraphics[width=\textwidth]{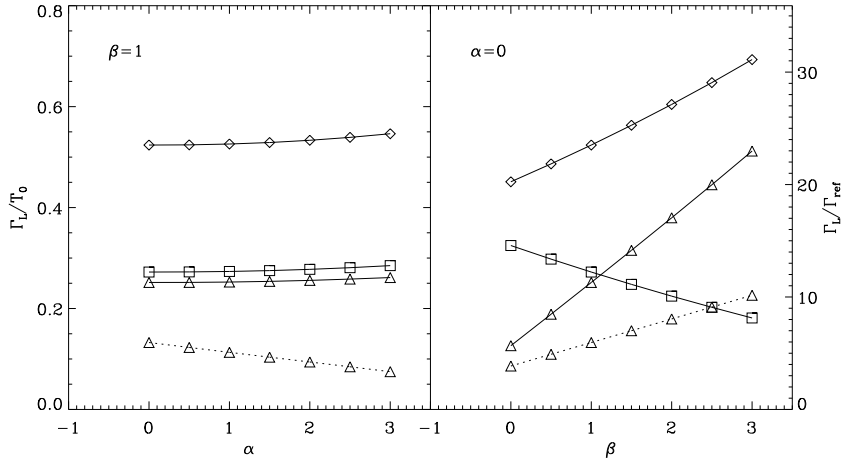}
  \caption{Total inner (squares) and outer (diamonds) torques (in
    absolute value) and differential Lindblad torque (triangles), for
    a vanishing smoothing length. The
    vertical scale is the same for both plots. On the left $y$-axis,
    the torque is normalized to $T_0=\pi q^2\Sigma
    a^4\Omega_p^2h^{-3}$, whereas on the right $y$-axis it is
    normalized to the reference value given by Eq.~(\ref{eq:9}). Both
    values differ by a factor $\pi/h$. The latter normalization is to
    be preferred, for it renders the differential Lindblad torque
    independent of the disk aspect ratio, if the disk is sufficiently
    thin. These plots should be compared to those of \citet{w97}
    (fig. 3) or those of \citet{mg2004} (fig. 2). The dotted line
    shows the differential Lindblad torque for a smoothing length
    $\epsilon=0.6H$.}
  \label{fig:comp}
\end{figure}

In Fig.~\ref{fig:comp} we present our results for disks with power law
profiles already studied in the literature in order to assess the
correctness of our implementation. We also display on these graphs the
differential Lindblad torque that we obtain with our fiducial
smoothing length.

\section{Calibration of the one-sided torque dependence on the wake's
  distance}
\label{sec:calibr-one-sided}
\begin{figure}
  \includegraphics[width=\textwidth]{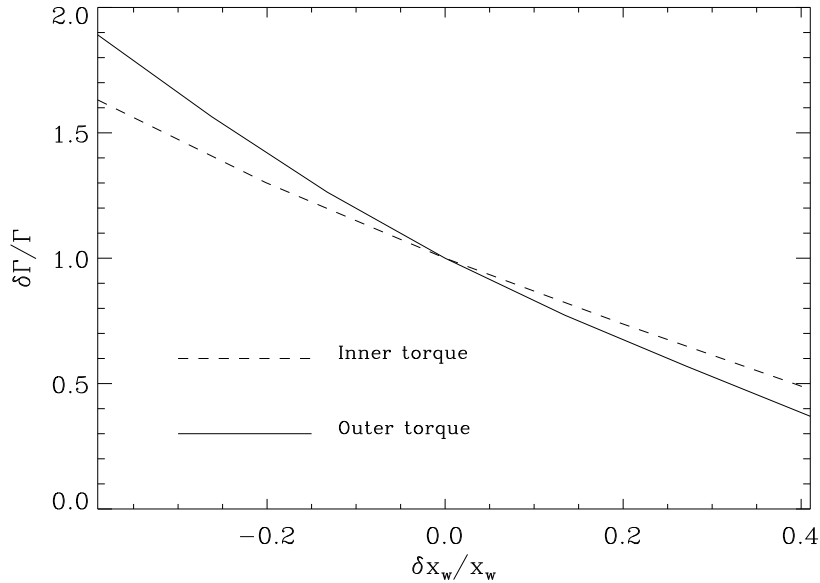}
  \caption{One-sided torque dependence on the distance to the wake.}
  \label{fig:ost_dist}
\end{figure}

We evaluate here how the one-sided torque varies as the wake recedes
or approaches the planet. We do this by means of numerical simulations
in which we maintain the planet on a fixed, uniform and  circular
orbit, but we vary its orbital frequency from run to run. Doing so
shifts the wake in a known amount. We measure
independently the inner and outer Lindblad torques, and we translate
the planet's frequency offset $\delta\Omega_p$ into a wake shift
$\delta x_w=-(2/3)a\delta\Omega_p/\Omega_p$.
The results are presented in Fig.~\ref{fig:ost_dist}. Not
surprisingly, the dependence is essentially similar for inner and outer
torques. We infer from these calculations that:
\begin{equation}
  \label{eq:94}
  \frac{\delta\Gamma}{\Gamma}\approx -2\frac{\delta x_w}{x_w},
\end{equation}
from the slopes measured in $\delta x_w=0$. The distance to the wake
$x_w$ is assumed to be $H$ \citep[see {\em e.g.} the torque density
distribution in][]{2010ApJ...724..730D}.

\end{document}